\RequirePackage{fix-cm}
\documentclass[smallextended]{svjour3}       % onecolumn (second format)
\smartqed  % flush right qed marks, e.g. at end of proof
\usepackage{graphicx}
\usepackage[dvipsnames]{xcolor}
\usepackage{array}
\usepackage{wrapfig}
\usepackage{multirow}
\usepackage{tabularx}
\usepackage{caption}
\usepackage{subcaption}
\usepackage{booktabs}
\usepackage{setspace}
\usepackage{tikz}
\usepackage{amssymb}
\usepackage{booktabs}
\usepackage{enumitem}
\usepackage{multirow}
\usepackage{longtable}
\usepackage{changepage}
\usepackage{hyperref}
\hypersetup{
    colorlinks=true,
    linkcolor=black,
    filecolor=black,      
    urlcolor=black,
    citecolor=black,
}
\usepackage{graphicx}

\usepackage{fancybox}
\usepackage{longtable}
\usepackage{tasks}
\usepackage{pdfpages}
\usepackage{fontawesome}
\journalname{Empirical Software Engineering}
\begin{document}

\title{The Impact of Personality on Requirements Engineering Activities: A Mixed-Methods Study
%\thanks{Grants or other notes
%about the article that should go on the front page should be
%placed here. General acknowledgments should be placed at the end of the article.}
}
%\subtitle{Do you have a subtitle?\\ If so, write it here}

\titlerunning{The Impact of Personality on RE}        % if too long for running head

\author{Dulaji Hidellaarachchi       \and John Grundy \and Rashina Hoda \and Ingo Mueller %etc.
}

%\authorrunning{Short form of author list} % if too long for running head

\institute{D. Hidellaarachchi\at
              Dept. of Software Systems and Cybersecurity, Monash University, Melbourne, Australia \\
              %Tel.: +123-45-678910\\
              %Fax: +123-45-678910\\
              \email{dulaji.hidellaarachchi@monash.edu}          %  \\
%             \emph{Present address:} of F. Author  %  if needed
           \and
           J. Grundy \at
              Dept. of Software Systems and Cybersecurity, Monash University, Melbourne, Australia \\
              \email{john.grundy@monash.edu}  
              \and
              R.Hoda\at
              Dept. of Software Systems and Cybersecurity, Monash University, Melbourne, Australia \\
              \email{rashina.hoda@monash.edu} 
              \and
              I.Mueller \at
              Dept. of Software Systems and Cybersecurity, Monash University, Melbourne, Australia \\
              \email{ingo.mueller@monash.edu}   
}

\date{Received: date / Accepted: 15 Nov 2023}
% The correct dates will be entered by the editor

\maketitle

\begin{abstract}
\textbf{Context:} Requirements engineering (RE) is an important part of Software Engineering (SE), consisting of various human-centric activities that require the frequent collaboration of a variety of roles. Prior research has shown that personality is one such human aspect that has a huge impact on the success of a software project. However, a limited number of empirical studies exist focusing on the impact of personality on RE activities. \textbf{Objective:} The objective of this study is to explore and identify the impact of personality on RE activities, provide a better understanding of these impacts, and provide guidance on how to better handle these impacts in RE. \textbf{Method:} We used a mixed-methods approach, including a personality test-based survey (50 participants) and an in-depth interview study (15 participants) with software practitioners from around the world involved in RE activities.  \textbf{Results:} Through personality test analysis, we found a majority of the practitioners have a high score on agreeableness and conscientiousness traits and an average score on extraversion and neuroticism traits. Through analysis of the interviews,  we found a range of impacts related to the personality traits of software practitioners, their team members, and external stakeholders. \textcolor{black}{It was found that having extraversion characteristics is considered as plus points compared to agreeableness, conscientiousness and openness to experience characteristics that have been stated as highly important to have when involved in RE activities. These impacts can vary depending on the RE activities, the overall software development process, and the people involved in these activities. Moreover, we found a set of strategies that can be helpful in overcoming some of the challenges associated with diverse personalities when involved in RE activities.} \textbf{Conclusion:} Our identified impacts of personality on RE activities and strategies serve to provide guidance to software practitioners on handling such possible personality impacts on RE activities and for researchers to investigate these impacts in greater depth in future. 

% \PACS{PACS code1 \and PACS code2 \and more}
%\subclass{MSC code1 \and MSC code2 \and more}
\end{abstract}

\section{Introduction} \label{section 1}

Personality is a human aspect with no one universally accepted definition. Although many theories have been developed related to personality, it is commonly referred to as \emph{individual differences} \cite{RN1556}. In this study, we use the definition by Mischel et. al. \cite{RN2976}, who define personality as ``\textit{a set of individual differences including personal habits, skills, memories, behaviours and social relationships that can be affected by the socio-cultural development of individuals}". 

Personality clashes or incompatibilities can affect the efficacy of collaboration and lead people to perform less effectively \cite{RN2977} \cite{RN2978} \cite{RN2979}. Investigating the impact of personality in software engineering (SE) has been an ongoing topic over many years \cite{CRUZ201594} \cite{RN1556} \cite{10.1145/2599990.2600012}. Various studies have examined the impact of the personality of software practitioners in the context of SE in general and for specific SE activities or contexts, such as development \cite{RN1620}, pair programming \cite{RN1616}, testing  \cite{10.5555/2819321.2819323}, software team composition \cite{RN2568}, team climate \cite{RN3008} to name a few.   A majority of these studies have focused on software practitioners involved in software development or testing in industrial settings. Some have used undergraduate/postgraduate students as participants \cite{RN2949} \cite{RN2577} \cite{ANVARI2017324}. Many studies were limited to particular organizations, countries, and geographic areas \cite{RN3008} \cite{CAPRETZ2015373}.
Based on the key findings of our previously conducted systematic literature review (SLR) on the impacts of human aspects on requirements engineering (RE)\cite{RN1600}, we identified personality as an important human aspect that needs to be further investigated in relation to RE. RE activities, such as eliciting, analysing, prioritizing, and managing software requirements play a vital role in SE and good requirements are  considered to be one of the most critical and challenging parts of SE. % which includes core activities such as requirements elicitation, analysis, documentation, validation and management \cite{RN2732} \cite{RN2973} \cite{RN2731}. 
Arguably, RE activities can be considered the most human-centric and socio-technically intensive activity in SE as they require extensive collaboration and understanding of the individuals involved \cite{ALSANOOSY20192394}. However, from the literature, we identified that there are very limited studies that directly focus on the personality of the software practitioners involved in RE activities \cite{RN1600}. 
Hence, we want to gain a more comprehensive understanding of how the personality of individuals influences RE activities in SE. Our broad research question is: \textbf{How does the personality of software practitioners influence requirements engineering activities?} 

\par \textcolor{black}{To answer this research question, we conducted a mixed methods exploratory study involving a personality test-based survey of 50 software practitioners involved in RE activities, followed by 15 in-depth interviews with those willing to discuss their experiences in-depth.} To gather their personality profiles, we used the standard IPIP NEO-120 Personality test based on the well-known five-factor model (FFM) of personality. We used the ranking scores defined in the IPIP NEO-120 test to analyse personality test data \cite{RN2997} and socio-technical grounded theory (STGT) \textit{for data analysis} to analyse the interview data \cite{hoda2021socio}. The main contributions of this study are as follows:
\begin{itemize}
    \item We identified some of the possible impacts of personality on RE activities related to software practitioners, their team members and their external stakeholders (customers/clients/end-users);
    \item We developed a set of guidelines for software professionals,  software teams and stakeholders, as well as academic and industry researchers, \textcolor{black}{who want to better understand the impact of personality on RE activities, including the challenges associated with diverse personalities and how they might go about overcoming some of the challenges; }and
    \item We identified a set of recommendations for future research into the impact of personality on RE activities.
\end{itemize}

\section{Background and Related Work} \label{section 2}

\subsection{Measuring Personality of Individuals}\label{section 2.1}
Numerous personality models have been formulated based on various personality theories to assess the personalities of individuals by characterizing human behaviours into a set of traits \cite{RN1556} \cite{RN2712}. The Five-Factor Model (FFM) is one of the most widely accepted personality models by psychologists and is now used in several SE studies on personality \cite{RN2956} \cite{RN1620}. The FFM integrates all personality characteristics into five main traits. These traits are \emph{Openness to experience, Conscientiousness, Extraversion, Agreeableness and Neuroticism}. These five traits represent one's personality at the broadest level of abstraction, and each trait summarizes a large number of distinct, more specific personality characteristics \cite{RN1618}.
The five main traits can be explained as follows \cite{RN1619}; 
\begin{itemize}
    \item[$\square$]{\textbf{Openness to Experience:} relates to individuals' intellectual, cultural or creative interests.  High-scored individuals with openness to experience tend to be imaginative, broad-minded and curious. In contrast, those at the opposite end of this spectrum usually show a lack of aesthetic sensibilities, favouring conservative values and preferring routine.}
\end{itemize}

\begin{itemize}
    \item[$\square$]{\textbf{Conscientiousness:} refers to individuals' focus on achievements. High-scored individuals tend to be hardworking, organized, able to complete tasks thoroughly on time, and reliable. Low-scored individuals on conscientiousness relate to negative traits such as being irresponsible, impulsive and disorganized.}
\end{itemize}

\begin{itemize}
    \item[$\square$]{\textbf{Extraversion:} relates to the degree of sociability, activeness, talkativeness, and assertiveness. A person is considered an extravert if they are friendly, comfortable in social relationships, active, assertive and outgoing. The opposite end of this spectrum shows a lack of social involvement, shyness, and prefers to be alone more than extraverted people. But, this does not mean that they are unfriendly or antisocial; rather, they are reserved in social situations.}
\end{itemize}

\begin{itemize}
    \item[$\square$]{\textbf{Agreeableness:} refers to positive traits such as cooperativeness, kindness, trust and warmth; agreeable individuals value getting along with others. Low-scored individuals on agreeableness tend to be sceptical, selfish and hostile.}
\end{itemize}

\begin{itemize}
    \item[$\square$]{\textbf{Neuroticism:} refers to the state of emotional stability of individuals. Low-scored individuals on neuroticism tend to be calm, confident and secure, whereas high-scored individuals on neuroticism tend to be moody, anxious, nervous and insecure.}
\end{itemize}

 \begin{figure}[t]
 %\centering
  \includegraphics[width=\linewidth]{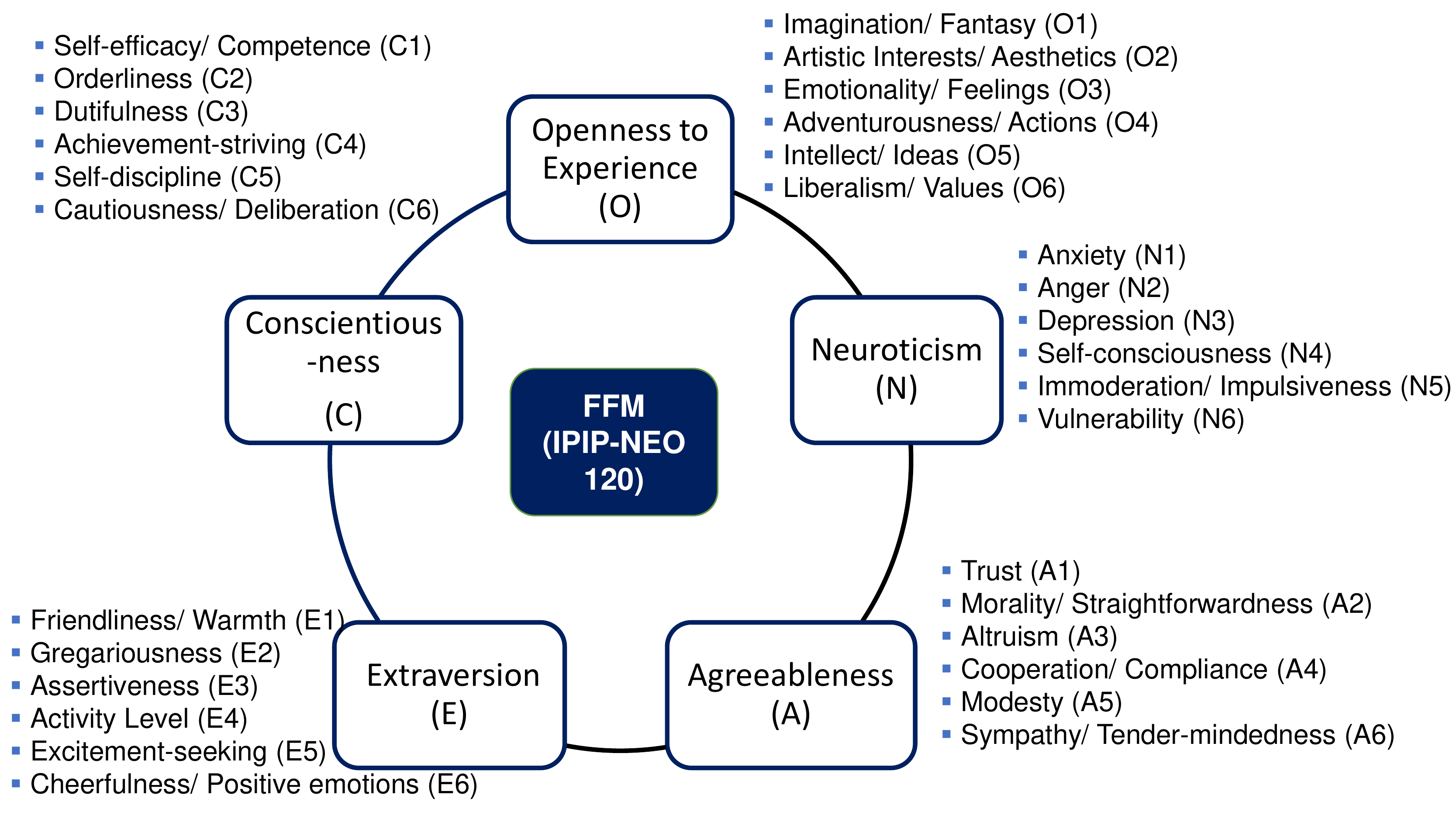}
  \caption{Five Dimensions and Thirty Facets of FFM (IPIP-NEO 120)}
  \label{Figure 01}
\end{figure}
These traits can be narrowed down to what is known as ``facets" which consist of six per trait (altogether 30 facets). The facets help to increase the precision and scope of FFM, thus enabling more accurate predictions \cite{RN1621}. Numerous instruments were developed to operationalize FFM. Among them, the International Personality Item Pool (\href{https://ipip.ori.org/}{\textbf{IPIP}}) is a freely available personality assessment instrument developed based on FFM. \textcolor{black}{\textbf{IPIP-NEO 120} is one of the short versions of the IPIP instrument and is designed to measure the five personality traits, including 30 facets, efficiently. This short version possesses acceptable measurement of reliability by evaluation of over 20,000 individual responses, and most recent studies focus on using the IPIP-NEO 120 version due to its acceptable reliability and practicality \cite{RN2997} \cite{RN1602} \cite{10.1145/3463274.3463327}. Goldberg's 300-item inventory is considered as the initial measure of IPIP, which has similar measuring constructs to those in NEO-PI-R \cite{RN2995}}.

\textcolor{black}{ The original IPIP-NEO version consists of 300 items, which will take 30-40 minutes to complete. Considering the convenience of use, shorter versions were created with less completion time. For example, the IPIP-NEO 120 inventory consists of 120 items and can be completed within 10-20 minutes, whereas the IPIP-NEO 50 inventory (50 items) can be completed within 3-8 minutes. The shortest version is the IPIP-NEO 20 inventory, and the IPIP-50 and IPIP-20 inventory only measure individuals' personalities related to five broad dimensions, not with 30 facets, making both inventories less reliable than others. Hence, we have used IPIP-NEO 120, the personality assessment instrument in our research study that is based on FFM (Figure \ref{Figure 01}) \cite{RN2997}.}

\subsection{Personality Research in Software Engineering}  \label{section 2.2}

%\todo{This section is also way too long - can you cut to ~1/2 page max?  Keep this detail for related work section in thesis...}

A considerable amount of studies have been conducted aiming to identify the impact of personality in the software engineering domain, and this significantly increased after 2002 \cite{CRUZ201594}. The systematic mapping study conducted by Cruz at al. \cite{CRUZ201594}, analysed 90 articles published between 1970 - 2010 and identified that 83\% of the research reported empirical findings on the role of personality in SE where pair programming and education were the most recurring research topics. Soomro et al. \cite{RN2712} conducted an SLR on the effect of the personalities of software engineers and how it is associated with team climate and team performance. The findings reported a relationship between software engineers' personalities and team performance and revealed that software team characteristics significantly impact software team performance. 

Another SLR was conducted by Barroso et al. \cite{RN2956}, focused on the influence of personality in SE, and found that most extracted primary studies focused on software designers' and programmers' personalities. 
%This study evaluated SE researchers' use of personality models and tests and how they identified the influence of personality on software engineers' work. 
%and diverse team climate compositions have been discussed mainly in terms of organizational behaviour and social science domains, but not much in the SE context. 
Salleh et al. \cite{RN1616} investigated the effect of personality traits in pair programming on higher education with five formal experiments. They identified that ``openness to experience" significantly differentiates paired students’ academic performance. An empirical study by Kanij et al. \cite{10.5555/2819321.2819323} investigated the impact of the personality of software testers by collecting personality profiles of 182 software practitioners. 45.1\% of them were software testers, and the rest were programmers. The results indicated that software testers obtained higher scores on the  ``conscientiousness" trait than other software practitioners. 
\par Xia et al. \cite{RN3005}, conducted a study with software professionals to identify the relationship between project managers' personalities and team personality composition and project success by investigating 28 completed software projects with 346 software professionals. %The DISC personality test was used and correlated the outcomes of the test with project success scores measured in six different dimensions: schedule, effort, risk, issue, quality, and customer satisfaction. Acua
The results indicated that project manager personality and team personality affect the success of software projects. They suggest focusing on relationships between personality and SE activities as their study only demonstrates the link between personality and overall project success. Kosti et al. \cite{RN2949} and Acua et al. \cite{RN2951} conducted empirical studies by collecting personality profiles of SE students. They identified significant relationships of personality with work preference and job satisfaction respectively. Further, they suggested conducting the research for real-world software teams.
%that considered significant associations between personality, and work preference in software engineering. They have collected personality data from 279 master level students, and the results indicated an association between personalities and work preference. 
 %In \cite{RN2951}, the relationships between personality, team processes and task characteristics, product quality and satisfaction in software development teams were analysed and found that teams with highest job satisfaction had the highest personality scores for agreeableness and conscientiousness personality traits. 35 student teams were incorporated in this study and suggested conducting the research for real-world software teams incorporating the teams' gender and age differences into it. 
Vishnubhotla et al. \cite{RN3008} investigated the relationship between personality and team climate focusing on software professionals in agile teams. 
%They have considered the FFM model for personality traits and the factors related to team climate within the context of agile teams working in the telecom company. 
The findings indicated a significant positive relationship between certain personality traits and team climate factors. 
%The study also suggested considering other human aspects besides personality traits to investigate the relationships related to team climate.
Mendes et al. \cite{RN1602} conducted a survey study with 63 software engineers investigating the relationship between decision-making style and personality within the context of software development. They identified seven statistically significant correlations between decision-making style and personality and built a regression model considering the decision-making style as the response variable and personality factors as independent variables. 

\par \textbf{Requirements Engineering} (RE)-related activities in SE are considered to be highly human-centred as they involve working with a diverse range of people such as stakeholders, software development team members, and other requirements engineers \cite{john2005human} \cite{RN2952} \cite{RN2445}. From our prior SLR study \cite{RN1600}, we identified that the impact of various human aspects needs to be studied more related to RE, and personality was identified as one such human aspect that has been considered often in SE studies, but RE has been considered as just one part of it or mainly limited to requirements elicitation \cite{RN2427} \cite{RN2568} \cite{RN2577} \cite{RN2942}. 
%Personality in RE was considered focusing on classifying effective personalities for web development limited to requirements elicitation revealing a relationship between human personalities and RE \cite{RN2430}.   
 %Their research revealed a relationship between human personalities and RE in web development and suggested conducting more research considering more human aspects and their impact on RE. 
 %Most of the systematic and empirical studies have focused on personality related to SE in general, or predominantly in software development, pair programming, testing, agile teams and global software development (GSD) context and the focus on RE has been mainly limited to requirements elicitation. 
Being motivated by our SLR, we surveyed 111 software practitioners involved in RE activities to better understand the perspective of human aspects, including personality. There we found the software practitioners' personality needs to be considered when they are involved in RE activities \cite{10.1145/3546943}. Having identified the importance of personality and its potential impact on RE as a key area worth investigating in our prior studies \cite{RN1600} \cite{10.1145/3546943}, acknowledging the research gap in this area motivated us to design and conduct this study to answer our above-mentioned research question.

\section{Research Methodology} \label{section 3}
This study aimed to understand participants’ perspectives on the impact of personality when involved in RE activities. 

%The quantitative approach was used to analyse closed-ended questions in the survey including the personality test assessment where we used standard IPIP NEO-120 personality test instrument to gather personality profiles of 50 software practitioners. Qualitative approach was used to analyse open-ended  for interview data analysis where we used Socio-Technical Grounded Theory (STGT) for data analysis \cite{hoda2021socio}.
\subsection{Study Design}
\begin{figure*}[b]
 \centering
  \includegraphics[width=\linewidth]{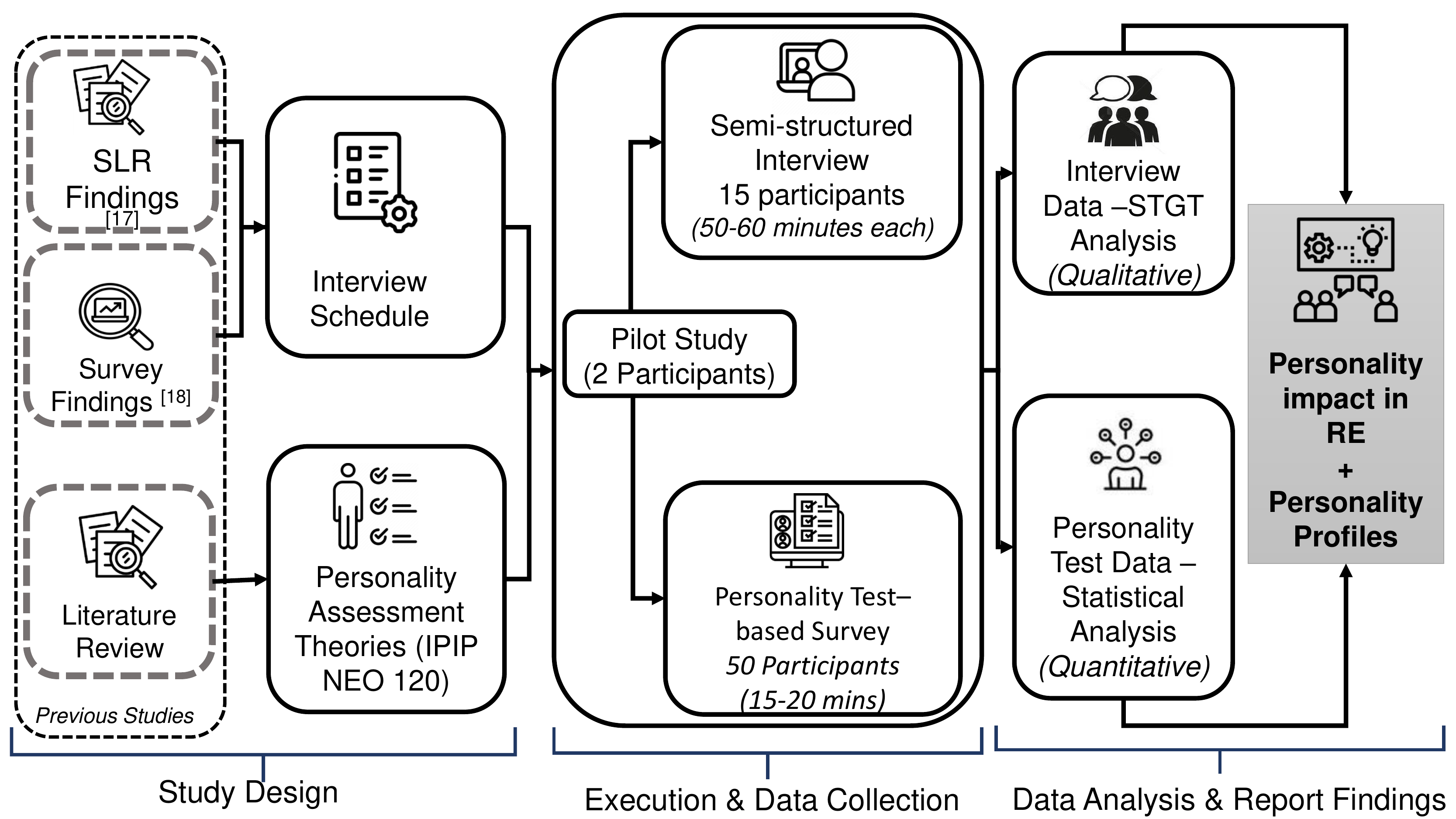}
  \caption{Overview of the mixed methods research design }
  \label{Figure02}
\end{figure*}

Figure \ref{Figure02} shows the design of our study. We applied a mixed-methods approach when conducting this study. Mixed here refers to both  qualitative and quantitative approaches being utilized. 
%and how they mitigate negative impacts (if any) based on their experiences.  
Our previous studies  \cite{RN1600} \cite{10.1145/3546943} identified personality as a human aspect that software practitioners believe greatly impacts RE activities. To obtain an in-depth understanding of the impact of personality on RE,  we designed a personality test-based survey and a \textcolor{black}{follow-up} interview study targeting software practitioners involved in RE activities. 
\subsubsection{Personality test-based survey}
The first step in our data collection was to obtain personality profiles of software practitioners involved in RE activities. \textcolor{black}{It includes a standard personality test, a set of demographic questions and an open-ended question to obtain the overall perspectives on the impact of personality on RE. By analysing the personality profiles, we aimed to obtain preliminary insights into the personalities of people doing RE activities, including whether there are any significant personality traits among the people doing RE activities}
%and to see if any particular personality trait/facets get mentioned a lot in the interviews referring to its' impact on RE. 
%\todo{repeats a bunch of text in Related Work - I suggest delete all that text there to reduce its size}
As the personality test, we used the standard IPIP-NEO 120 test, a self-assessment quantitative instrument, to obtain the personality profiles of our participants (Section \ref{section 2.1}).  
%which was developed based on the most scientifically validated personality assessment model called FFM.
It consists of 120 items/statements, and participants are required to indicate how each statement best describes them. For example, the first statement is ``\emph{I worry about things"}, where participants should indicate how much it related to themselves via a Likert scale from ``very inaccurate" to ``very accurate". Apart from the personality test, we included several demographic questions such as participants' age range, gender, country of residence, education qualifications, and work experience to identify their involvement in RE activities. We collected their employment details such as job role/title, a summary of their main job responsibilities, how often they are involved in RE activities, and their domain and software development methods. An open-ended question was used to collect their overall opinion on the impact of personality on RE activities. We used the \emph{Qualtrics} platform to design and distribute the survey (Appendix \ref{A}). 

\subsubsection{Interview study}
\textcolor{black}{Along with the personality test-based survey, we conducted a \textcolor{black}{follow-up} interview study to get an in-depth understanding of the impact of personality on RE activities. In this study, our main focus was to elicit software practitioners' perceptions of personality and its impact on RE activities referring to themselves, their team members and their external stakeholders.}
%The second stage of our data collection used an interview-based study to get an in-depth understanding of the impact of personality on RE activities.
To do this, we designed a semi-structured interview schedule/ guide consisting of three sections. %The set of questions contained both open-ended and close-ended types of questions. 
%Our focus was on the influence of personality on RE activities in general and specific to five personality dimensions in the FFM model. 
The first section collected detailed demographic information about the participants, including their involvement in RE activities. The second section collected the participant's views on the influence of personality in RE activities. The views on the influence of their personality, their team members' personalities and their external stakeholders (customers/ clients/ end users) personalities were focused on related to RE activities and \textcolor{black}{how they overcome some of the challenges that occurred due to diverse personalities (if any) }when involved in RE activities. The last section asked participants'  opinions on the impact of other human aspects besides personality on RE activities (if any) based on their experiences. Our semi-structured interview schedule can be found in Appendix \ref{B}. 

\subsubsection{Pilot study}
After designing the personality test-based survey and the interview study, we conducted a pilot study with two software practitioners from our professional networks to validate the clarity and understandability of the questions, the time reported to complete the studies and to get their suggestions on improving both studies.  Both provided feedback on the personality test-based survey and interview study, where we modified the questions to improve the clarity based on their suggestions, \textcolor{black}{except for the standard 120 statements in the personality test} and finalized the survey and interview studies. 

\subsection{Data Collection}
\begin{table}[]
\centering
\caption{\centering Participants' Demographics Information}
\label{TABLE: Participants' demographics}
\resizebox{\columnwidth}{!}{%
\footnotesize
\begin{tabular}{@{}ll@{}}
\toprule
\multicolumn{1}{l}{\emph{Demographic Information} }      & \emph{Participants \% }                                                \\ \midrule
\multicolumn{1}{l}{\textbf{Age Range}}                                                        \\ \midrule
21-30    & 36\%   \\
31-40     &  40\%   \\
41-50   &  10\%         \\
Above 50                &  14\%        \\
 \midrule
\multicolumn{1}{l}{\textbf{ Job roles of the Participants}}   \\ \midrule
Software Engineer    & 40\%       \\
IT project Manager    & 12\%    \\  
Lead Business Analyst & 10\%    \\
Tech Lead & 6\%  \\
\begin{tabular}[c]{@{}l@{}}Business Analyst, ERP \& Digital Solution Architect, User\\ experience Designer, Senior Consultant (IT), Software QA \\Engineer, Junior Systems Engineer  \end{tabular}  & 4\% each  \\
 \begin{tabular}[c]{@{}l@{}} Product Owner, Software Testing Manager, IT Development\\ \& Re-engineering Leader, Lead Human-centered Designer\end{tabular}  & 2\% each  \\ \midrule
\multicolumn{1}{l}{\textbf{Countries of the Participants}}   \\ \midrule
United Kingdom                                           & 20\%   \\
South Africa    & 12\%   \\
Australia, Sri Lanka  & 8\% each      \\
India, Poland, Spain           & 6\%           \\
\begin{tabular}[c]{@{}l@{}}Ireland, New Zealand, Netherlands, \\Portugal, Hungary, Italy, Mexico \end{tabular}                                         & 4\% each  \\  
\begin{tabular}[c]{@{}l@{}} France, Indonesia, Nepal \end{tabular}                                         & 2\% each  \\ \midrule
\multicolumn{1}{l}{\textbf{Work Experience in SE}}   \\ \midrule
Less than 1 year      & 4\%   \\
Between 1-5 years   & 34\%   \\
Between 5-10 years  & 34\%      \\
More than 10 years   & 28\%           \\

\bottomrule
\end{tabular}%
}
\end{table}
%\todo{Can you cut any of this back a bit too:}
The target population of our study was software practitioners involved in RE activities \textcolor{black}{and we collected personality profiles from 50 software practitioners to analyse in our study. After obtaining the required ethics approval\footnote{Monash Ethics Review Manager (ERM) reference number: 29072}, \textcolor{black}{we advertised our personality test-based survey on social media (LinkedIn, Twitter), within our own professional network and the \textbf{Prolific} platform \cite{PALAN201822}.}
\textcolor{black}{Prolific is a data collection platform dedicated to academic purposes with over 130,000 active users and widely used in software engineering research to recruit participants \cite{russo2022anecdote} \cite{cucolacs2023impact} \cite{russo2021developers}. }
We used the built-in options in the prolific platform to filter participants based on our target audience. We applied the filter options for participants' employment status as ``full-time", employment sector as ``information technology", and industry as ``software" and specifically mentioned that we are looking for software practitioners involved in RE activities and \textcolor{black}{applied questions to filter participants who are involved in RE activities to obtain our target participants.} We recruited 35 participants via Prolific. Each of them was rewarded 8.72 AUD after completing the survey.}

\textcolor{black}{Our interview study was also advertised on social media (LinkedIn, Twitter) and within our own professional networks along with the personality test-based survey. Participation was voluntary, and we recruited 15 practitioners with experience in RE activities worldwide.  As a first step, they completed the personality test-based survey which took around 15-20 minutes to complete, making a total of 50 personality profiles. }Then, the semi-structured interviews were carried out, with each interview lasting 50-60 minutes. Due to the pandemic, all the interviews were conducted online and were audio recorded. Table \ref{TABLE: Participants' demographics} shows the detailed demographic information of the participants.

 %Secondly, Qualitative data were collected through semi-structured interviews. The interviewed software practitioners were employees of private or governmental organizations in their respective countries, having between 3 to 26+ years of experience in doing RE activities. 
 
%\par We wanted to compare the personality traits of those involved in RE activities. We collected a total of 50 personality profiles  from software practitioners involved in RE activities. 

%As we had obtained only 15 personality profiles of the software practitioners interviewed,  This resulted in a total of 50 personality profiles to analyse.

\subsection{Data Analysis}

Both qualitative and quantitative data were collected in our study. 
%Therefore, we used a mixed methods approach for the data analysis. 
\textbf{Quantitative data analysis} (personality test data) was carried out with the data collected through the personality test-based survey. We followed the standard personality test analysis method mentioned in IPIP-NEO 120 personality test (\href{https://ipip.ori.org/}{\textbf{IPIP}}) \footnote{https://ipip.ori.org/} to analyze the personality test data via Microsoft Excel. \textcolor{black}{As mentioned in Section \ref{section 2.1}, the personality test consists of 120 items/statements and each item is related to one personality trait and one personality facet simultaneously. Therefore, each statement in the personality test represents one personality facet which is defined under one of the five personality traits in the FFM model. For example, the statement \emph{``I make friends easily"} relates to the ``friendliness" facet under the extraversion personality trait.  Further, each item is defined as + key or - key. A + keyed item adds value in ascending order, from 1 (very inaccurate) to 5 (very accurate) points to the personality score for their respective personality trait and facet, whereas - keyed items have inverted the scores (1 - very accurate and 5 -very inaccurate) in the personality test. }

\begin{table}[b]
\centering
\caption{Example of the scoring method of the personality test as defined in IPIP-NEO 120 test}
\label{TABLE: example of personality test}
\resizebox{\columnwidth}{!}{%
\footnotesize
\begin{tabular}{@{}llll@{}}
\toprule
{\textbf{\begin{tabular}[c]{@{}l@{}}Neuroticism (17+7=24)\end{tabular}}}                                                       
& \textbf{\begin{tabular}[c]{@{}l@{}}\end{tabular}}

& \textbf{\begin{tabular}[c]{@{}l@{}} INT01 Answers\end{tabular}} 
& \textbf{\begin{tabular}[c]{@{}l@{}} Item Scores\end{tabular}}      

\\ \midrule
{\textbf{\begin{tabular}[c]{@{}l@{}}Anxiety (N1)\end{tabular}}}                                                       
& {\begin{tabular}[c]{@{}l@{}}\end{tabular}}

& {\begin{tabular}[c]{@{}l@{}}\end{tabular}} 
& {\begin{tabular}[c]{@{}l@{}} \end{tabular}}\\

 \multirow{4}{*}{\begin{tabular}[c]{@{}l@{}}  + Keyed \end{tabular}}
 & Worry about things & Very accurate & 5\\
 & Fear for the worst & Very accurate & 5 \\
 &Am afraid of many things & Very accurate & 5\\
 & Get stressed out easily & Very accurate & 5 \\ \midrule
 
 {\textbf{\begin{tabular}[c]{@{}l@{}}Anger Hostility/ Anger (N2)\end{tabular}}}                                                       
& {\begin{tabular}[c]{@{}l@{}}\end{tabular}}

& {\begin{tabular}[c]{@{}l@{}}\end{tabular}} 
& {\begin{tabular}[c]{@{}l@{}} \end{tabular}}\\

 \multirow{3}{*}{\begin{tabular}[c]{@{}l@{}}  + Keyed \end{tabular}}
 & Get angry easily & Neither accurate nor inaccurate & 3\\
 & Get irritated easily  & Moderately accurate & 4 \\
 & Lose my temper & Moderately inaccurate & 2\\
- keyed & Am not easily annoyed & Moderately inaccurate & 4 \\

\bottomrule
\end{tabular}%
}
\end{table}

\textcolor{black}{Table \ref{TABLE: example of personality test} shows an example of the scoring method used in the personality test analysis as defined in the IPIP-NEO 120 test with two facets of the neuroticism trait from participant INT01. IPIP-NEO 120 test is designed such that the total number of questionnaire items (statements) per trait is 24 (4 items per facet, and there are six facets under one personality trait); when calculated, each personality trait score is between 24 and 120. This is because, for each item, the minimum and maximum scores vary between 1 to 5 as of the defined scores, irrespective of whether it's a + key or - key item. Similarly, each personality facet's score is between 4 and 20 (as the total number of questionnaire items per facet is 4). The personality trait and facets scores are presented as percentages, and based on the percentages, the results are categorized as ``low", ``average," and ``high" according to whether the score is approximately in the lowest 30\%, middle 40\% or highest 30\% of the scores, assisting to identify the level of each personality trait of an individual. The results obtained via analysing the personality profiles are discussed in section \ref{4.2}, and an example of a personality profile can be seen in Appendix \ref{C}.}
 %\begin{figure*}[t]
 %\centering
 % \includegraphics[width=\linewidth]{personalityscores.png}
%  \caption{Example of the scoring method of the personality test as defined in IPIP-NEO 120 test}
%  \label{personality scores}
%\end{figure*}

\begin{figure*}[b]
 \centering
  \includegraphics[width=\linewidth]{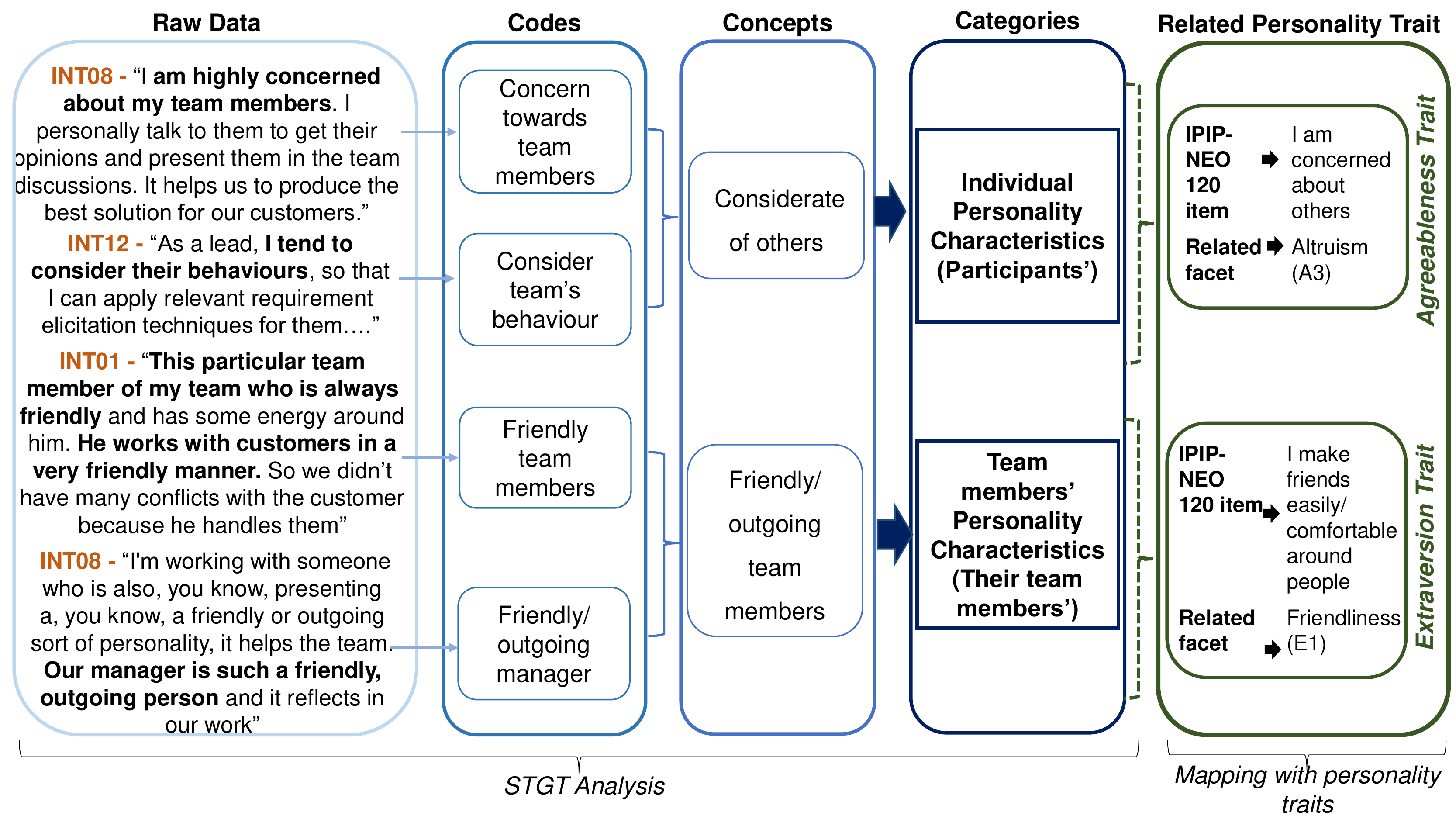}
  \caption{Example of the STGT analysis, mapped to personality traits and facets}
  \label{STGT example}
\end{figure*}

\par \textbf{Qualitative data analysis} was carried out on interview data. We used  \textbf{socio-technical grounded theory (STGT)} \emph{for data analysis} \cite{hoda2021socio}, \textcolor{black}{the basic stage of the STGT method }\textcolor{black}{which is particularly suited for analyzing open-ended data and understanding the insights in socio-technical contexts such as ours, i.e. studying the impact of personality in RE activities.} We transcribed the interview sessions using (\href{https://otter.ai/}{\textbf{otter.ai}}) \footnote{https://otter.ai/}, an online transcription tool, with the consent of the participants and stored \& analyzed the data using NViVO. The data collection and analysis were iterative, where we conducted ten interviews in the first iteration and five in the second iteration (Table \ref{TABLE: Interview participants' demographics}). \textcolor{black}{As shown in Figure \ref{STGT example}, we followed the open coding approach to generate concepts and categories with constant comparison and memoing techniques in STGT. We grouped similar codes to define various concepts, and then these concepts were identified and grouped into categories.  With the concepts \& categorization and considering the definitions of each personality trait and facet from the IPIP-NEO 120 personality test, we were able to map these concepts and categories into related personality traits and facets (Table \ref{TABLE 4: Individual personality}}). Furthermore, we wrote \textbf{memos} to record key insights while following the open coding activities. Below is an example of a memo that we recorded related to the impact of the neuroticism dimension on RE activities. We have discussed these insights generated from memoing in section \ref{section 5}. \\

\cornersize{.05} 
\ovalbox{\begin{minipage}{10cm}{
\small
\par \textbf{Memo on ``Impact of neuroticism trait on RE activities"}: The majority of the interview participants (n = 12 out of 15) mentioned that the neuroticism characteristics of the people negatively impact RE. They mentioned that when people involved in RE activities are \emph{``anxious", ``insecure", ``overthinking", ``moody", and ``reserved"}, it negatively impacts RE activities, specifically in requirements elicitation. All these adjectives emphasize the neuroticism trait, indicating that it negatively impacts RE activities. Interestingly, many of them mentioned these related to their team members or customers' characteristics and did not particularly mention their own neuroticism characteristics. E.g., INT01, a lead business analyst, mentioned that their team members' `` anxious" and ``overthinking" behaviour creates confusion between customers and the team in requirements elicitation, resulting in getting incomplete requirements during the project. However, they tend to explain the impact related to their team members' characteristics rather than their own characteristics, which also can occur due to the nature of their individual personalities. Hence, the relationship between software practitioners' personalities and their team members'/customers' personalities can also be further investigated related to RE/SE.}
\end{minipage}} \\

The results obtained via STGT analysis are presented in section \ref{4.3}, which mainly discusses the impact of personality in RE activities and in section \ref{4.4}, which provides \textcolor{black}{a set of strategies to overcome some of the challenges occurred due to diverse personalities} when involved in RE activities. \textcolor{black}{We also discuss the findings from the qualitative data analysis in light of the results of the personality profiles in section \ref{qualVquant}.}

\section{Findings} \label{section 4}

\subsection{Demographics of the participants}

\begin{table*}[]
\centering
\caption{\centering Demographics of the Interview Participants (*P-ID: Participant ID, *Ex. in SE: Experience in software industry) }
\label{TABLE: Interview participants' demographics}
\resizebox{\textwidth}{!}{%
\begin{tabular}{@{}llllllll@{}}
\toprule
\multicolumn{1}{l}{\textbf{\begin{tabular}[c]{@{}l@{}}P-ID*\end{tabular} } }      & \textbf{\begin{tabular}[c]{@{}l@{}} Age \\Range\end{tabular}}     & \textbf{Country}     & \textbf{\begin{tabular}[c]{@{}l@{}} *Ex.in\\ SE \\(yrs)\end{tabular}}     &\textbf{\begin{tabular}[c]{@{}l@{}}Ex.in \\RE\\ (yrs)\end{tabular}}  &\textbf{\begin{tabular}[c]{@{}l@{}} Job Role/\\Title\end{tabular} }   & \textbf{Project Domains}    & \textbf{\begin{tabular}[c]{@{}l@{}}S/W Dev\\ Methods \end{tabular}}                         \\ \midrule

INT01 & 21-30  & Sri Lanka & 3.5 & 3.5 &  \begin{tabular}[c]{@{}l@{}}  Lead Business \\Analyst \end{tabular}  & \begin{tabular}[c]{@{}l@{}} IoT \& \\Telecom\\-munication \end{tabular}  & Agile (Scrum) \\  \\

INT02 & 21-30  & Sri Lanka & 6+ & 6+ & \begin{tabular}[c]{@{}l@{}} Business Analyst \\\& Project Manager \end{tabular} & \begin{tabular}[c]{@{}l@{}}  Transport \\\& Logistics \end{tabular} & Agile (scrum) \\  \\

INT03 & Above 50  & Indonesia & 24 & 18 & \begin{tabular}[c]{@{}l@{}}IT Development \\ \& Re-engineering \\Leader \end{tabular} & \begin{tabular}[c]{@{}l@{}} Manufacturing, \\Web, ERP  \end{tabular} & \begin{tabular}[c]{@{}l@{}} Agile \& \\Waterfall  \end{tabular}\\ \\

INT04 & 31-40  & Sri Lanka & 4 & 4 & Business Analyst & Finance, IT & Agile (scrum) \\  \\

INT05 & 31-40  & Netherlands & 11 & 5+ & \begin{tabular}[c]{@{}l@{}}Software Engineer \\ \& Application \\Consultant \end{tabular} & \begin{tabular}[c]{@{}l@{}} Telecom\\-munication \end{tabular}& Agile (scrum)\\  \\

INT06 & 31-40  & New Zealand & 10 & 10 & \begin{tabular}[c]{@{}l@{}} Technical \\Team Lead \end{tabular} & \begin{tabular}[c]{@{}l@{}}  Finance, Health,\\ Insurance\end{tabular} & \begin{tabular}[c]{@{}l@{}}  Agile \\(2 weeks \\sprints)\end{tabular}\\  \\

INT07 & 21-30  & Nepal & 5 & 2 & \begin{tabular}[c]{@{}l@{}} Software Engineer\\ (IoT, DevOps) \end{tabular} & \begin{tabular}[c]{@{}l@{}}  Transport \& \\Logistics\end{tabular} & Agile (scrum)\\  \\

INT08 & 21-30  & Australia & 5 & 2 & Software Engineer & Real estate & Agile (scrum) \\  \\

INT09 & Above 50  & Australia & 26 & 26 & \begin{tabular}[c]{@{}l@{}}Lead Human-\\centered Designer\\/ Information \\Architecture\end{tabular} & \begin{tabular}[c]{@{}l@{}} Transport \& \\Logistics\end{tabular} & \begin{tabular}[c]{@{}l@{}} Agile \& \\Waterfall  \end{tabular}\\   \\

INT10 & 21-30  & India & 9 & 5 & Business Analyst & Utility services & Agile (scrum) \\  \\ 

INT11 & Above 50  & New Zealand & 20+ & 20+ &  \begin{tabular}[c]{@{}l@{}}  Lead Business \\Analyst \end{tabular} & Health & Agile (scrum)\\  \\

INT12 & Above 50  & Australia & 26+ & 26+ &  \begin{tabular}[c]{@{}l@{}}  Senior Project \\Manager \end{tabular} & Health, Insurance & Agile (scrum)\\  \\

INT13 & 21-30  & Sri Lanka & 5 & 5 & \begin{tabular}[c]{@{}l@{}}Lead Business \\ Analyst, \\Scrum Master \end{tabular} & Health & Agile (scrum)\\   \\
INT14 & 31-40  & India & 9 & 6 &  \begin{tabular}[c]{@{}l@{}}  Lead Business \\Analyst \end{tabular} &  Health & Agile (scrum)\\ \\
INT15 & 31-40  & India & 6+ & 5 & \begin{tabular}[c]{@{}l@{}}  Senior Software\\ Engineer (SSE)\end{tabular}  &  \begin{tabular}[c]{@{}l@{}} IoT \& \\Telecom\\-munication\end{tabular}  & Agile (scrum) \\ 

\bottomrule
\end{tabular}%
}
\end{table*}

Our study consists of a personality test-based survey (50 software practitioners) and an interview study (15 software practitioners). The majority of the participants were male (66\%), with ages ranging between 31 to 40 years (40\%). The most common roles are software engineer (40\%), followed by IT project manager (12\%) and business analyst lead (10\%), where the majority of the participants (60\%) possess a bachelor's degree in computer science or software engineering.  Table \ref{TABLE: Participants' demographics} summarises all the demographic information of the participants. The majority of the participants were from the United Kingdom (20\%), followed by South Africa (12\%), Australia (8\%) and Sri Lanka (8\%). An equal number of participants (34\%) have 1-5 years and 5-10 years of work experience in the software industry, whereas 28\% have more than ten years of work experience.  When specifically considering interview participants (15 software practitioners), their work experience in the software industry expands from 3 to 26+ years. Most of them have been fully involved in RE activities throughout these years.  

Table \ref{TABLE: Interview participants' demographics} provides detailed demographic information about the interview participants, where 08 out of 15 participants are male. Most participants (68\%) use agile software development methods, such as Scrum, Kanban, and XP, and 20\% use both agile and structured (e.g. waterfall) software development methods. The rest of the 12\% only use structured  (e.g waterfall) software development methods.  Among the majority who use agile software development, 55.8\% use scrum, and the rest use other methods such as kanban, XP and lean software development. The participants have experience in a variety of project domains where the majority of them (28\%) were finance domain, followed by health (24\%), transport \& logistics (18\%), IT (10\%), government services (8\%), manufacturing (8\%), education (6\%), real estate (4\%), IoT \& telecommunication (4\%), utility services (4\%), and insurance (2\%).

\par Our target participants were software practitioners involved in RE activities. We confirmed this by asking participants to rate how often they were involved in major RE activities when they completed the personality test-based survey. An equal percentage of participants (38\% each) indicated that their involvement in RE activities, such as eliciting/analyzing/prioritizing/ managing software requirements, is almost every day or a couple of times a week. 20\% of the remaining participants mentioned that they were involved in these RE activities a couple of times a month. In contrast, only 4\% of the participants mentioned their involvement in RE activities is rare. We also provided a set of major RE activities to identify further how much the participants were involved in RE as a part of their job. 

 \begin{figure}[t]
 %\centering
  \includegraphics[width=\linewidth]{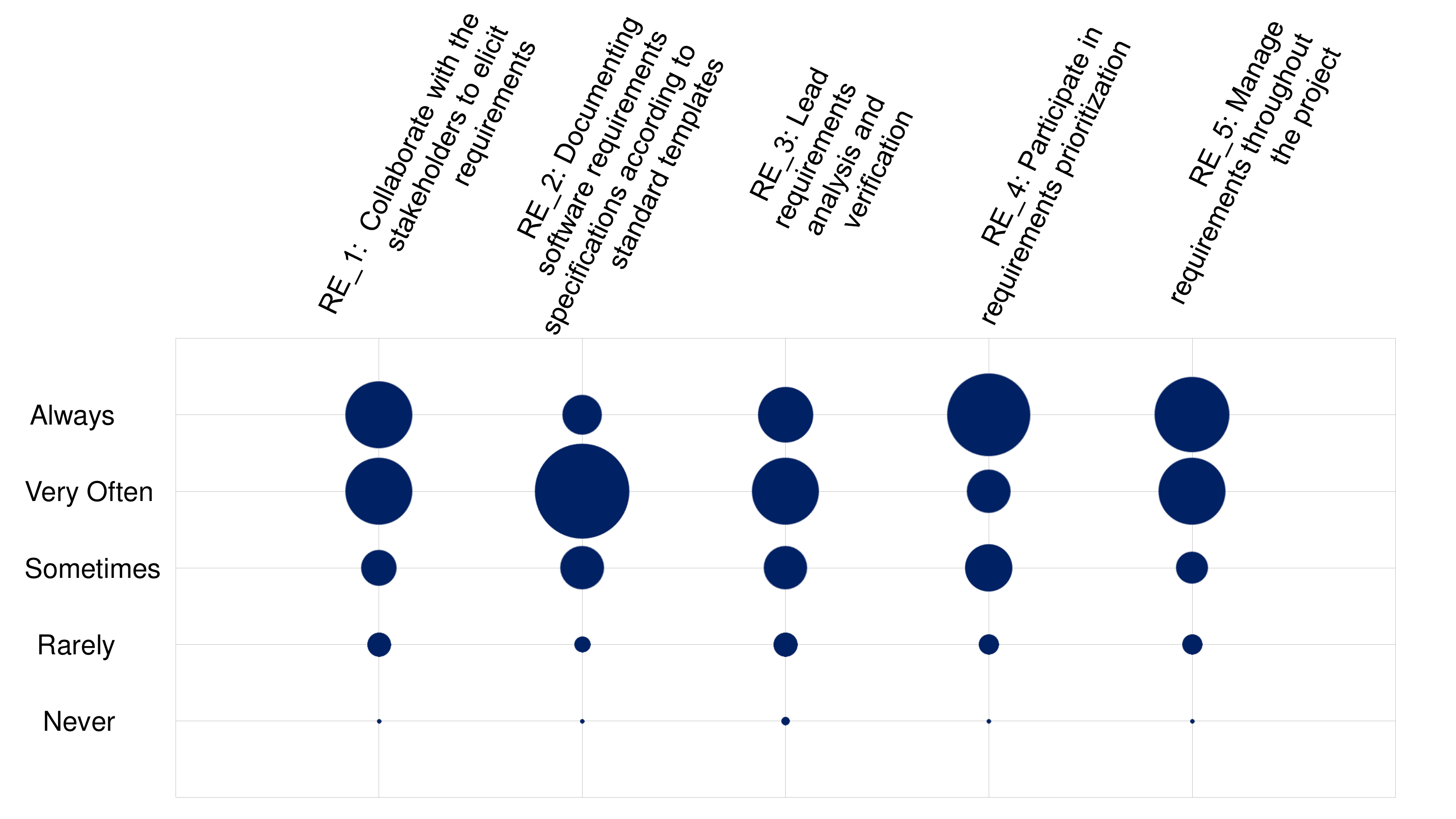}
  \caption{Participants' involvement in RE activities}
  \label{Figure 03}
\end{figure}

\par As shown in figure  \ref{Figure 03}, the majority of the participants are either always or very often involved in these RE activities. Among them, the majority (72\%) are always or very often involved in requirements management throughout the project.  68\% of participants either always or very often collaborate with stakeholders to elicit requirements and to document them in software requirements specifications. 64\% of the participants always or very often participate in requirements prioritization, whereas 62\% always or very often lead requirements analysis and verification. The least amount of participants mentioned that they were never involved in these given RE activities, which is a maximum of 4\% for activity `` lead requirements analysis \& verification". For all the other RE activities given, only 2\% of participants mentioned that they were never involved in those activities. We also asked them to mention other job responsibilities they are involved in, where 10\% are involved in the implementation and 4\% in testing/bug fixing, architecture designing and creating user guides after feature development as the tasks apart from the given list of activities. It shows that we were able to recruit our target participant group for this research study.

\subsection{Personality Characteristics of the practitioners involved in RE activities} \label{4.2}

We collected 50 personality profiles from software practitioners involved in RE activities. We used the standard IPIP-NEO 120 personality assessment test, designed based on FFM, as our personality test (section 2.1). It includes 120 items/statements that help to analyze a person's personality, referring to five broad dimensions and 30 facets (figure \ref{Figure 01}).  A sample personality profile of a participant can be seen in Appendix C. 
 \begin{figure}[]
 %\centering
  \includegraphics[width=\linewidth]{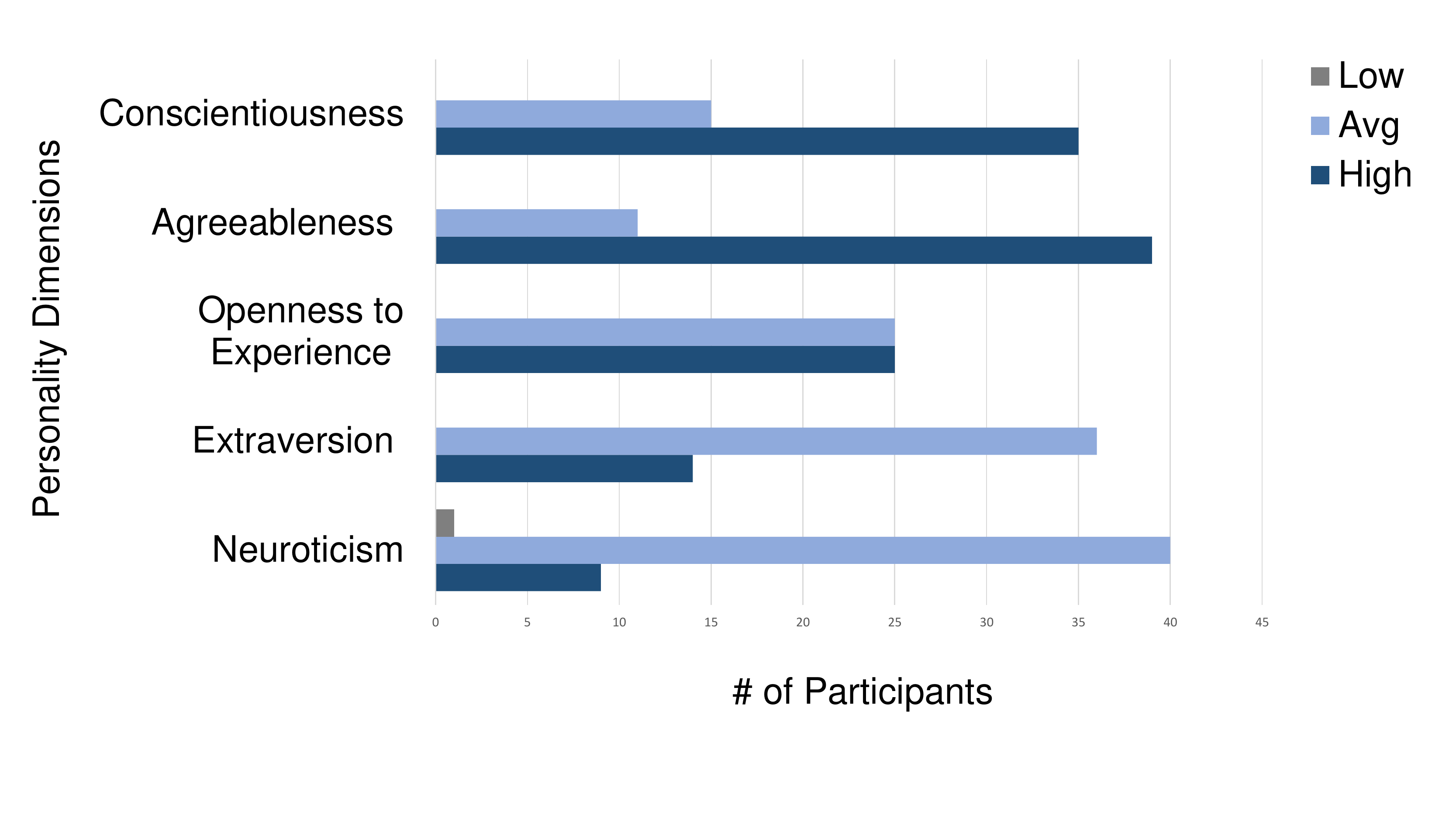}
  \caption{Variation of Personality traits/dimensions of the participants}
  \label{Figure 04}
\end{figure}
By analyzing the 50 personality profiles that we have collected; we identified that although every individual possesses a mix of personality traits, some personality traits are more significant than others. As shown in Figure  \ref{Figure 04}, agreeableness (78\%) and conscientiousness (70\%) are identified as significant (high-scored)  personality traits among the majority of our participants. An equal number of participants (50\% each) have a high and average score related to the openness to experience trait. The majority of the participants have an average score in their extraversion and neuroticism traits, where 72\% are average in extraversion and 80\% are average in neuroticism.
The `high', `average' and `low' categories depicted in Figure \ref{Figure 04} are decided based on each individual's scores through analyzing their personality tests. We identified that the majority of the participants have high agreeableness and conscientiousness traits and average extraversion and neuroticism traits. None of the participants is low in agreeableness, conscientiousness, openness to experience and extraversion traits, and only one participant (2\%) is identified as low in neuroticism. 

\begin{figure}[t]
 %\centering
  \includegraphics[width=\linewidth]{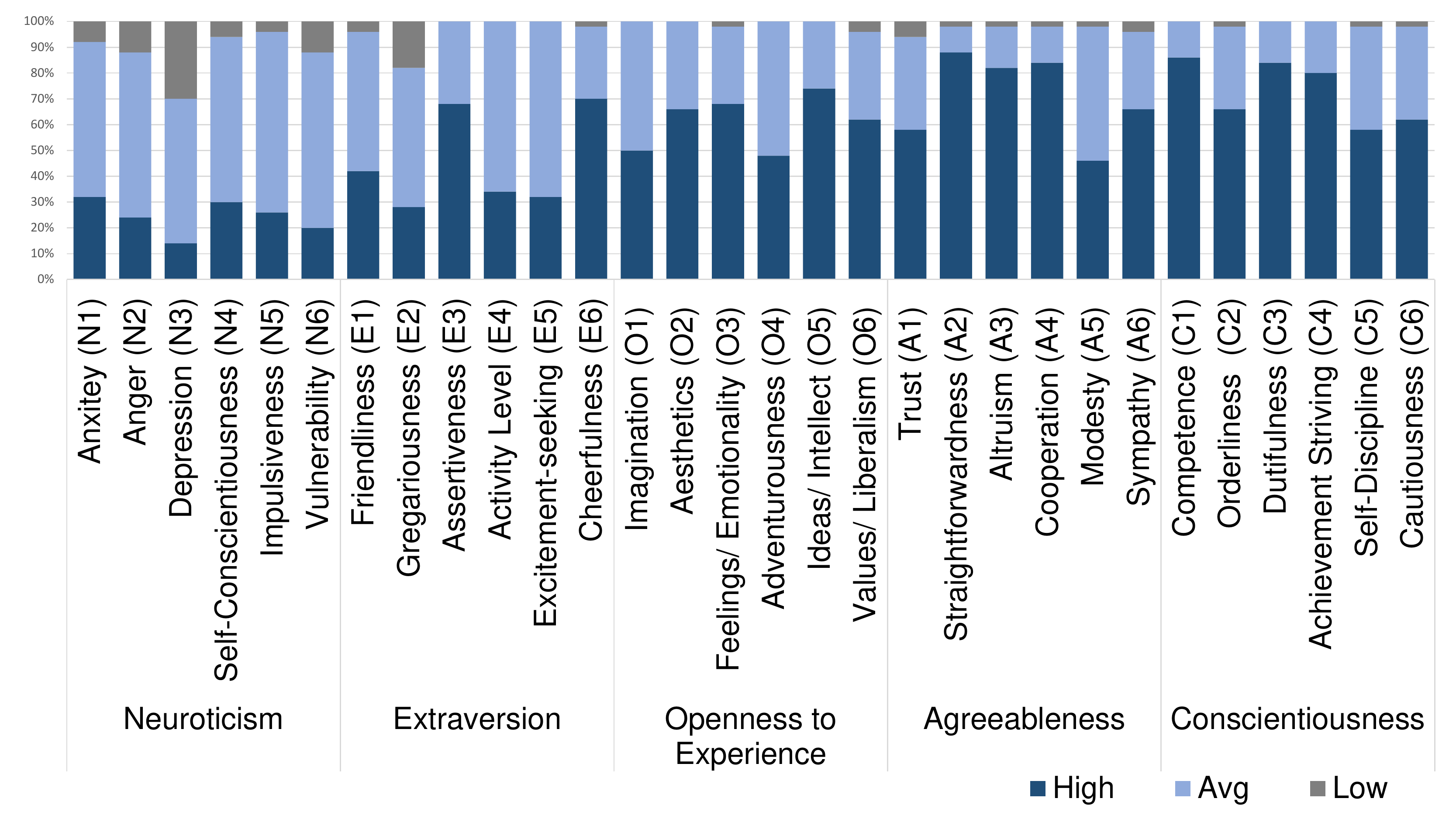}
  \caption{Variation of the facets of the Personality traits/dimensions of the participants}
  \label{Figure 05}
\end{figure}

 Each personality trait consists of six facets that can be used to describe each personality trait in detail. Hence, an individual's personality can be described with five personality traits and thirty facets. As shown in figure  \ref{Figure 05}, most of the facets related to agreeableness, conscientiousness and openness to experience are significantly high among participants. In contrast, the majority of the facets related to extraversion and neuroticism have average scores. Considering the agreeableness trait, the most common facet among the participants, its' straightforwardness (88\%), cooperation (84\%), and altruism (82\%) facets are significantly high in the majority of the participants compared to sympathy, trust and modesty facets. This shows that the participants' personalities related to agreeableness can be best described as straightforward, cooperative and altruistic individuals. 
\par Considering conscientiousness, participants obtained high scores in competence (86\%), dutifulness (84\%) and achievement striving (80\%) facets, making it the second highest personality trait that can be seen among the participants. 
However, compared to facets of the other three personality traits, most of the facets of agreeableness and conscientiousness have high scores among participants, indicating that these two personality traits are the most common traits among the participants.
Among the facets of openness to experience, the trait that has an equal percentage of high and average scores, ideas/intellect (74\%), feelings/emotionality (68\%), aesthetics (66\%), and values/liberalism (62\%) facets have high scores compared to the imagination and adventurousness facets. 
\par However, none of the facets of openness to experience reaches the level of scores of agreeableness and conscientiousness. This indicates that participants tend to have more openness to experience behaviours than extraversion and neuroticism, but not as high as agreeableness or conscientiousness behaviours. Referring to the facets of the extraversion trait, only cheerfulness (70\%) and assertiveness (68\%) are considerably high among participants. In contrast, friendliness, gregariousness, activity level and excitement-seeking can be seen as average-scored facets among the participants. Neuroticism is the trait where most facets have average scores among the participants, indicating that neuroticism behaviour is mainly average among participants. Among their average neuroticism behaviour, the majority of the participants seem to be impulsive (70\%), vulnerable (68\%), angry (64\%) and self-conscientious (64\%). Compared with other personality traits, it can be identified that a considerable amount of participants are low in neuroticism facets such as depression (30\%), vulnerability (12\%), anger (12\%), and anxiety (8\%). In contrast, the majority of the others facets are hardly seen in a low category. By analysing these 50 personality profiles, we identified that the participants tend to have a high-score in agreeableness, conscientiousness and openness to experience than in extroversion and neuroticism. 

\subsection{Impact of personality in RE activities} \label{4.3}

\textcolor{black}{While personality tests allowed us to understand the significant personality traits of our participants, we needed to understand their perspectives on how the personality of the individuals impacts RE activities and how they overcome some of the challenges (if any) that occurred due to the different personalities of the people involved in RE activities. So, we conducted semi-structured follow-up interviews with 15 software practitioners who are predominantly involved in RE activities to gain deeper insights into the impact of personality in RE activities.} Their demographics are summarised in Table \ref{TABLE: Interview participants' demographics}. Throughout this section, we include several original quotes from the interviews to illustrate our findings. Almost every participant mentioned their view on the impact of personality on RE activities. \textcolor{black}{The majority of participants agree that there is a positive influence or challenges that occurred due to diverse personalities in RE and the importance of identifying the impact and how to overcome some of the challenges}. Only 3 participants claimed there was little to no impact of personality, mainly referring to clients/ customers/ end users' personalities \faCommentsO\emph{``I think you can gather requirements from end users irrespective of their personality" - INT09 [lead human-centred designer/information architecture]} rather than those of team members or their own.

\par Most of our interview participants experienced the impact of personality when doing RE activities irrespective of the people involved -- their own personality, team members' personalities, or clients/ customers/ end users' personalities even in a virtual environment:
\faCommentsO \emph{``Even in these unprecedented times, where all the things are virtual, and you do not get a chance to work with people at the office, still the various personalities of the people actually, really important to conduct our work" - INT05 [software engineer/application consultant]}.
 
\par Referring to the impact of personality on RE activities, \textcolor{black}{ we identified that the impact could enhance or be challenging on completing the RE activities, the outcomes of RE activities,} the overall outcome of the project, and/or the performance of the team and people involved. For example, according to INT05,  a software engineer \& application consultant, personality highly impacts the most important RE activity: \faCommentsO``\emph{my personal opinion in the requirement engineering, the personality mainly impacts on the requirements gathering part, which is the most important part, not the other side.. giving required output.."} INT13, a lead business analyst, explained that personality impacts the individual performance as well as the performance of the team:  \faCommentsO\emph{``personality makes a huge difference for a person as well as for the team because when they are involved in requirements engineering activities with various personalities, that affects a lot for you and the people you're working with to do your best. Eventually, it improves our performance"}. Through in-depth analysis of the interviews, we have identified that these personality impacts can be further explained as related to software practitioners' personalities, their team members' personalities and their clients/customers/end users' personalities.\\

\begin{itemize}
    \item \textbf{Impacts related to software practitioners' individual personality }
\end{itemize}
Participants identified various personality characteristics and how these characteristics impact them when they are involved in RE activities. Table \ref{TABLE 4: Individual personality} summarizes these participant-reported impacts, relating impact to individual personalities and highlighting \textcolor{black}{whether it is \textbf{enhancing} (\faArrowCircleUp) or \textbf{challenging} (\faArrowCircleDown) when involved in RE activities.} We have also identified that these characteristics can directly or indirectly be mapped to the five personality traits/dimensions and their facets. Some important aspects are discussed as follows:

\par \faArrowCircleUp\hspace{0.1cm}\textbf{Ability to `get things done':} The majority of the participants (13) mentioned that their ability to 'get things done' has a large impact when involved in RE activities. Individuals with this characteristic know how to get things done to achieve what they need to. \textcolor{black}{They do this by prioritizing their work when necessary while working with their team members and customers.} This personality facet characteristic helps them to work with different customers during requirements elicitation in order to obtain the actual software system requirements. It also helps them in their dealings with internal team members (other developers and project managers) when translating specifications into technical tasks to reduce future system complications.  \\
For example: \faCommentsO\emph{``so now I am dealing with him in a very different way. I make things simple for him. I only asked him the key questions first and then gradually gathered requirements with smaller discussions"- INT12 [Senior project manager].}\\
\faCommentsO\emph{``But When you take it positively, when you know how to get the thing right at the first phase, you can reduce that number of errors, bugs, complications on whatever that is to come with the dev team when giving specifications as technical tasks"- INT13 [Lead BA/Scrum master].}

\par \faArrowCircleUp \hspace{0.1cm}\textbf{Hardworking nature:} 6 participants mentioned that \textcolor{black}{their hardworking nature personality facet enhances how they are involved in RE activities and its outcomes}. Through this hardworking nature, they can provide the best solution to their customers, resulting in high customer satisfaction and completing their work tasks within the scheduled time, saving budget and time. It also results in delivering reliable work quickly and highlights that they can identify more requirements via their hardworking nature, leading them to dig deeper into the relevant domain. 
\faCommentsO\emph{``I will be doing more research on the area and then get back to them with some sort of a plan, then carry out the work with the team. In the end, they are very happy about the work" - INT02 [BA/ Project manager].}\\
The \textbf{above two characteristics} can be directly mapped with the \textbf{conscientiousness} trait, where the ``ability to get things done" is related to the competence facet and ``hardworking nature" is related to the achievement-striving facet. 

\par \faArrowCircleUp\hspace{0.1cm}\textbf{Considerate of others:} 14 participants said that their personality facet of concern towards others involved in RE/SE-related activities impacts them when they are involved in RE activities. Consideration towards other team members leads them to focus on their allocated tasks and solve conflicts between customers and the team when involved in RE activities. Moreover, it helps to gather opinions from everyone, which leads us to get the best solution among them. \faCommentsO\emph{``I am highly concerned about my team members, I personally talk to them to get their opinions and present them in the team discussions. It helps us to produce the best solution for our customers" -INT08 [Software engineer].}\\
Concern towards their customers helps them better understand their behaviours and change the requirements elicitation techniques accordingly. For example, if the customer tends to be more reserved and is not expressive of what they need, changing the requirement elicitation technique will help get the best outcome:
\faCommentsO\emph{``As a lead, I tend to consider their behaviours so that I can apply relevant requirement elicitation techniques for them. For example, we had a very reserved client, not talking much, so we changed our way of collecting requirements, and it was a success. We only asked main points in the meeting, and when we needed more information, we emailed him"  - INT12 [Senior project manager].}

\begin{table*}[]
\caption{Impact of personality related to software practitioner's individual personality}
\label{TABLE 4: Individual personality}
\resizebox{\textwidth}{!}{%
\begin{tabular}{@{}llll@{}}
\toprule
 \textbf{\begin{tabular}[c]{@{}l@{}} Charac\\-teristic \end{tabular}}  & \textbf{Impact on RE/SE} & \textbf{Related Facet} & \textbf{\begin{tabular}[c]{@{}l@{}} Personality \\ \\Trait\end{tabular}}\\ \midrule
\begin{tabular}[c]{@{}l@{}} Ability to\\ get things\\ done \end{tabular} & \begin{tabular}[c]{@{}l@{}} (\faArrowUp) Handing various customers during requirements elicitation\\ (\faArrowUp) Get the actual requirements \\ (\faArrowUp) Ability to deal internal team members (developers) when translating \\specifications to technical tasks\\ (\faArrowUp) Reduce future complications/errors/bugs \\ (\faArrowUp) Solving issues in requirement changes
\end{tabular}
& \begin{tabular}[c]{@{}l@{}} Competence/ \\Self-efficacy \\(C1)	\end{tabular}
&\begin{tabular}[c]{@{}l@{}} Conscien\\-tiousness \end{tabular}\\ \\  

\begin{tabular}[c]{@{}l@{}} Hard-working\\ nature \end{tabular}
& \begin{tabular}[c]{@{}l@{}} (\faArrowUp) Ability to deliver reliable work quickly\\ (\faArrowUp) Providing the best solution to the customers, result\\ in high customer satisfaction\\ (\faArrowUp) Completion of tasks within the timeline saving budget \& time \\ (\faArrowUp) Getting more requirements via digging deeper in the domain	\end{tabular}
&  \begin{tabular}[c]{@{}l@{}}  Achievement-\\ striving \\(C4) \end{tabular}& \begin{tabular}[c]{@{}l@{}} Conscien\\-tiousness \end{tabular}\\ \\     

\begin{tabular}[c]{@{}l@{}} Being \\Systematic \end{tabular} & \begin{tabular}[c]{@{}l@{}}(\faArrowUp) Following a periodic plan helpful in allocate tasks in the team \\ (\faArrowUp) Being methodological leads successful task completion \\ (\faArrowUp) Working with a plan leads to not missing out requirements   \end{tabular}     & Self-discipline (C5) & \begin{tabular}[c]{@{}l@{}} Conscien\\-tiousness \end{tabular}\\ \\  

\begin{tabular}[c]{@{}l@{}} Desire for \\successful\\ task completion \end{tabular} &   \begin{tabular}[c]{@{}l@{}}(\faArrowUp) Successful elicitation \& analysis of requirements \\ (\faArrowUp) Ability to complete tasks within deadlines \\ (\faArrowUp) Increase the accuracy of the work 
      \end{tabular}                    					       
& \begin{tabular}[c]{@{}l@{}} Competence/ \\Self-efficacy \\(C1)	\end{tabular}
& \begin{tabular}[c]{@{}l@{}} Conscien\\-tiousness \end{tabular}\\ \\ 

\begin{tabular}[c]{@{}l@{}}   Considerate \\of others \end{tabular}  &  \begin{tabular}[c]{@{}l@{}}   (\faArrowUp) Lead team members to focus on their allocated task \\ (\faArrowUp) Solve conflicts between customers and the team members \\ (\faArrowUp) Help to gather everyone's opinion, leading towards getting the\\ best solution \\ (\faArrowUp) Help to understand customer behaviours and change the elicitation \\techniques accordingly     \end{tabular}  & Altruism (A3) & \begin{tabular}[c]{@{}l@{}} Agreea\\-bleness \end{tabular} \\\\

 \begin{tabular}[c]{@{}l@{}}  Cooperative \\nature    \end{tabular}  &
   \begin{tabular}[c]{@{}l@{}}  (\faArrowUp) Helpful in requirements specification and prioritization \\ (\faArrowUp) Helpful in managing requirement changes \\ (\faArrowUp) Make it easier to get the team and the customers into one page \end{tabular}   &  
   \begin{tabular}[c]{@{}l@{}} Compliance/ \\Cooperation\\ (A4) \end{tabular} & \begin{tabular}[c]{@{}l@{}} Agreea\\-bleness \end{tabular} \\\\ 
   
 \begin{tabular}[c]{@{}l@{}} Trusting \\others \end{tabular} &                                                 \begin{tabular}[c]{@{}l@{}} (\faArrowUp) Trusting the team members will lead to improve their work performance \\ (\faArrowUp)  High-level of trust ensure strong collaboration with customers, \\result in getting complete requirements \\ (\faArrowUp)  Trusting the development team, result in getting the best possible \\solution for customers \end{tabular}   & Trust (A1) & \begin{tabular}[c]{@{}l@{}} Agreea\\-bleness \end{tabular} \\ \\
 
  \begin{tabular}[c]{@{}l@{}}    Helpfulness      \end{tabular} &  \begin{tabular}[c]{@{}l@{}}     (\faArrowUp) Lead to solve difficulties in the design/ solution with ease \\ (\faArrowUp) Getting more expert inputs to deal with last minute change requests \\ (\faArrowUp) Improve the overall success of the project, resulting in successful project\\ execution \end{tabular} & Altruism (A3) & \begin{tabular}[c]{@{}l@{}} Agreea\\-bleness \end{tabular} \\\\
  
  \begin{tabular}[c]{@{}l@{}}    Enjoy\\ in-depth \\conversations \end{tabular}  &  \begin{tabular}[c]{@{}l@{}}     (\faArrowUp) Having in-depth discussions with customers lead to get actual requirements \\ (\faArrowUp) Connects the team with the customers from the ground level, leading \\to having less conflicts in later part of the project \\ (\faArrowUp) Helpful to know in-and-out of the project \\ (\faArrowUp) Ability to handle change requests from the development team     \end{tabular}  &  \begin{tabular}[c]{@{}l@{}} Ideas/ \\Intellect\\ (O5) \end{tabular}& 
  \begin{tabular}[c]{@{}l@{}} Openness to\\ Experience  \end{tabular}\\ \\
  
   \begin{tabular}[c]{@{}l@{}}   Willingness \\to try \\new things       \end{tabular} & 
     \begin{tabular}[c]{@{}l@{}}  (\faArrowUp) Enhance the individual learning about the project \\ (\faArrowUp) Improve the design of the project with advance functionalities \\ (\faArrowUp) Increase customer satisfaction with new approaches         \end{tabular} &    \begin{tabular}[c]{@{}l@{}} Actions/ Adven-\\turousness \\(O4)  \end{tabular} 
     & \begin{tabular}[c]{@{}l@{}} Openness to\\ Experience  \end{tabular}  \\\\
 
    \begin{tabular}[c]{@{}l@{}} Imaginative\\ nature \end{tabular} &  \begin{tabular}[c]{@{}l@{}}  (\faArrowUp) Helpful in foreseen what kind of model/interface best suit for collected\\ requirements \\ (\faArrowUp)  Increase the creativity of the final outcome of the project \\ (\faArrowUp) Ability to identify hidden/ unexpressed requirements by the customers   \\ (\faArrowUp) Ability to identify interactions between collected requirements   \end{tabular}
    & \begin{tabular}[c]{@{}l@{}}  Fantasy/ \\Imagination \\(O1) \end{tabular} & \begin{tabular}[c]{@{}l@{}} Openness to\\ Experience  \end{tabular}  \\\\

 \begin{tabular}[c]{@{}l@{}}   Ability to \\interact with \\various roles       \end{tabular} &                                                 \begin{tabular}[c]{@{}l@{}} (\faArrowUp) Initiate and continue to have successful conversations with the customers \\ (\faArrowUp) Create a positive connection with customers, result in getting\\ customer support throughout the project \\ (\faArrowUp) Getting more requirements via approaching the different level of customers \\ (\faArrowUp) Ability to get the required inputs and outputs from various involved parties \\of the project \\ (\faArrowUp) Increase the easiness of working within the team    \end{tabular} & \begin{tabular}[c]{@{}l@{}}    Warmth/ \\Friendliness\\ (E1)\\ Gregariousness \\(E2)  \end{tabular} & Extraversion \\\\
 
 \begin{tabular}[c]{@{}l@{}}Prefer to \\control/\\take the lead:    \end{tabular} & 
 \begin{tabular}[c]{@{}l@{}}  (\faArrowUp) Handling difficult customers with direct involvement to identify and solve\\ issues, reducing the conflicts with the team \\ (\faArrowUp) Helpful in maintaining the continuity of the project without \\having interruptions\\ for final output \\ (\faArrowUp) Ability to ensure that everyone in the team get a voice \\ (\faArrowDown) Over-controlling can reduce the team performance \\ (\faArrowDown) Reduce the others' interest  towards the project, result in completing tasks \\just to get by   \end{tabular}
 & \begin{tabular}[c]{@{}l@{}} Assertiveness\\ (E3)\end{tabular}  & Extraversion \\\\
 
  \begin{tabular}[c]{@{}l@{}}  Being curious        \end{tabular} &
 \begin{tabular}[c]{@{}l@{}}    (\faArrowUp) Able to do a good job with gathering requirements \\ (\faArrowUp) Ability to gather detailed requirements \\ (\faArrowUp)  Curious mindset increases the individuals' level of involvement in each task     \end{tabular} & \begin{tabular}[c]{@{}l@{}}  Excitement-\\Seeking \\(E5) \end{tabular} & Extraversion \\\\
 
 \begin{tabular}[c]{@{}l@{}}     Getting anxious     \end{tabular} &
 \begin{tabular}[c]{@{}l@{}}    (\faArrowDown) Create confusion between customers and the internal team \\ (\faArrowDown) Inability to understand what customers actually need due to anxiousness \\ (\faArrowDown) Inability to complete tasks on time, affecting the individual and team\\ performance   \end{tabular} & Anxiety (N1) & Neuroticism \\\\
 
  \begin{tabular}[c]{@{}l@{}}   Insecure feeling       \end{tabular} &
 \begin{tabular}[c]{@{}l@{}} (\faArrowDown) Direct impact on individual and team performance when involved \\in RE/SE tasks \\ (\faArrowDown) Increase the conflicts between team members \\ (\faArrowDown) Reluctant to stepping out from conform zone, reducing the chances to learn        \end{tabular} & \begin{tabular}[c]{@{}l@{}} Vulnerability \\(N6)  \end{tabular}& Neuroticism \\
  \\ \bottomrule 
\end{tabular}}%
\end{table*}

\par \faArrowCircleUp\hspace{0.1cm}\textbf{Cooperative nature:}  The majority of the participants (13 out of 15) highlighted that their personality facet of a cooperative nature is beneficial when they are involved in RE activities, both within their team and with their customers. Cooperativeness greatly impacts requirements specification, prioritization, and handling of requirement changes, taking the team and the clients to one page:   
\faCommentsO\emph{``It did not help to do requirement engineering unless you have qualities such as cooperativeness, it is really helpful when we prioritize the requirements and handling requirement changes in later phase" - INT04 [Business Analyst].}\\
These characteristics can be directly mapped with the \textbf{agreeableness} dimension, where the ``considerate of others" can be referred to as altruism and ``cooperative nature" can be directly referred to as compliance/ cooperation. 

\par \faArrowCircleUp\hspace{0.1cm}\textbf{Enjoy in-depth conversations:} Individuals who have a personality facet of enjoy having in-depth conversations said \textcolor{black}{that it enhances how they carry out RE activities.} It was mentioned that without having in-depth discussions with the customers, getting actual requirements is difficult. In-depth discussions connect the team with the customers from the ground level, leading to fewer conflicts in the later part of the project. It is also helpful to know in and out of the project, which will assist in handling change requests from the development team: 
\faCommentsO\emph{``if you want to know, understand the system they want, then you have to have a long fruitful conversation with the clients, you can't just sit and expect them to give you the requirement. I always enjoy having long conversations with them, and then I know what they actually want" - INT10 [Business Analyst].}

\par \faArrowCircleUp\hspace{0.1cm}\textbf{Willingness to try new things:} Of 15, 12 participants mentioned that they are always open to trying new things, and their personality facet of having an exploratory nature has a high impact when they are involved in RE activities. They said it enhances the individual learning about the projects, improves the project's design with advanced functionalities and increases customer satisfaction with new approaches, which will benefit them:  
\faCommentsO\emph{``I am always open to try new things, especially when designing each project. Mostly, it has positively affected to my team and customers, they are very satisfied with new functionalities" - INT14 [Business Analyst].}\\
Both of these characteristics indicate that individuals with the \textbf{openness to experience} trait positively impact RE activities where the ``enjoy in-depth conversations" can be mapped with ideas/ intellect facet and ``willingness to try new things" can be mapped with actions/ adventurous facet.

\par \faArrowCircleUp\hspace{0.1cm}\textbf{Ability to interact with various roles:} 9 participants mentioned that their personality facet of an ability to interact with various people/roles greatly benefits them when involved in RE activities. Interactions with internal and external parties are required. The ability to interact with various people greatly impacts the success of a particular task. For example, individuals with high interaction skills are needed to initiate conversations with customers, which will be helpful in having a positive connection with the customers and result in getting good support from them throughout the project.  
\faCommentsO\emph{``It is needed when you start conversations with your clients, creates some kind of bond with them. If you have a good bond with your clients, they will support you with everything. So having high interaction skills, I think, it is like one of the strong skills that you need as in personality" - INT04 [Business Analyst]}

\par \faArrowCircleUp /\faArrowCircleDown\hspace{0.1cm}\textbf{Prefer to control/take the lead:} It was mentioned that individuals' personality facet of a preference to take control or take the lead has a mixed impact when they are involved in RE activities (10 out of 15). For example, when dealing with a difficult customer who is having issues with the team, it is important to take the lead and intermediate with them. This is done by directly contacting the customers to identify issues and provide immediate solutions required to maintain the continuation of the project without interruptions for the final output. It also ensures that everyone in the team gets a voice: 
\faCommentsO\emph{``I will directly contact them and interact with them. Try to identify the problem and make sure it will not impact our work and the final project" - INT03 [IT development \& re-engineering lead].}\\
However, \textcolor{black}{we identified that this personality facet sometimes creates challenges for developers when carrying out RE activities.} For example, if there is an over-controlling lead in the team, it will directly impact the team's performance. It reduces team members' interest in the project, and they start to complete the tasks just to get by, which will eventually impact the project's outcome, leading to customer dissatisfaction: \faCommentsO\emph{``However, if you try to over-control them, it will definitely impact as others will just do the tasks, and it will be clearly seen in their performance" - INT06 [Technical team lead].}\\
These characteristics can be mapped onto the \textbf{extraversion} dimension where the ``ability to interact with various roles" can be referred to as friendliness and gregariousness facets, and ``prefer to control/ take the lead" can be referred to as the assertiveness facet.

\par \faArrowCircleDown\hspace{0.1cm}\textbf{Getting anxious:} \textcolor{black}{Individuals with a personality facet of getting anxious tend to have a challenging impact when they are involved in RE activities.} Individuals' confidence level, deadlines, and reluctant to change make them anxious, resulting in creating confusion between customers and the internal team, not understanding what customers actually need and inability to complete tasks on time, which will eventually affect their performance:   
\faCommentsO\emph{``I know personally, within my experience, I know that there are times when I'm just anxious, because,  probably, I'm not confident about the work, maybe I think I've not done my homework well, so it does affect my performance" - INT10 [Business Analyst].}

\par \faArrowCircleDown\hspace{0.1cm}\textbf{Insecure feeling:} \textcolor{black}{Individuals' personality facet of having insecure feelings also creates a challenging impact when they are involved in RE activities.} A feeling of insecurity comes from their self-doubts, seeing other team members' performance, or making changes to their allocated tasks. It will also directly impact their performance and the team's performance, increasing the conflicts between team members. Overall, it will impact the outcome of the project: 
\faCommentsO\emph{``A few years back, I was kind of insecure because I have self-doubt of myself, seeing all my colleagues working as pros, so you tend to feel insecure, and it really impacts my work. I couldn't give what the team is asking from me" - INT07 [Software engineer].}\\
Both of these characteristics can be directly mapped onto the \textbf{neuroticism} dimension, where the majority of the participants mentioned that it has a challenging impact when doing RE activities. The ``getting anxious" characteristic can be directly referred to anxiety facet, whereas the ``insecure feeling" can be referred to ``vulnerability" facet.\\

\begin{itemize}
    \item \textbf{Impacts related to software practitioners' team members' personality }
\end{itemize}
Almost every participant provided insights on how the personality of their team members impacts them when they are involved in RE activities. \textcolor{black}{This included how the team members' diverse personalities enhance or create challenges when involved in RE activities, the overall outcome of the project and their team performance. We have identified several personality characteristics that are helpful in completing RE activities that every participant wants to see within their team members and some characteristics that may create challenging situations when involved in RE activities}.  Table \ref{TABLE 5: Team members personality} summarizes the set of important characteristics, mapping with respective personality dimensions/traits and facets accordingly and discussed as follows;

\begin{table*}[]
\caption{Impact of personality related to software practitioners' team members' personality}
\label{TABLE 5: Team members personality}
\resizebox{\textwidth}{!}{%
\begin{tabular}{@{}llll@{}}
\toprule
 \textbf{\begin{tabular}[c]{@{}l@{}} Charac\\-teristic\end{tabular} }  & \textbf{Impact on RE/SE} & \textbf{Related Facet} & \textbf{ \begin{tabular}[c]{@{}l@{}} Personality \\ Trait \end{tabular}}\\ \midrule
\begin{tabular}[c]{@{}l@{}} Friendly/ \\outgoing \\nature  \end{tabular} & \begin{tabular}[c]{@{}l@{}} (\faArrowUp) Initiate and conduct the discussions with customers \\ (\faArrowUp) Helpful in getting actual, in-depth requirements  \\ (\faArrowUp) Helpful in dealing with constant requirements changes, unrealistic\\ \& impractical requirements by customers  \\ (\faArrowUp) Reduce the conflicts between the team and the customers \\ (\faArrowUp) Keeping the energy of the team and carry forward the teamwork \\ (\faArrowDown) Reduce customer involvement in RE due to not getting enough\\ opportunity to express their opinions \\ (\faArrowDown) Getting incomplete requirements, creating requirements gaps
\end{tabular}
& 	\begin{tabular}[c]{@{}l@{}} Warmth/ \\Friendliness \\(E1)\end{tabular}
& Extraversion\\ \\    

\begin{tabular}[c]{@{}l@{}} Leading \& \\ Controlling \\nature\end{tabular} &
\begin{tabular}[c]{@{}l@{}} (\faArrowUp) Help to manage the internal team \& lead the whole team \\ (\faArrowDown) Other team members loss of interest towards the project \\ (\faArrowDown) Internal team conflicts \\ (\faArrowDown) Not getting best outcome due to over-power others' opinions\end{tabular} & \begin{tabular}[c]{@{}l@{}} Assertiveness\\ (E3) \end{tabular}& Extraversion \\ \\

\begin{tabular}[c]{@{}l@{}} Curious\\ nature \end{tabular} & \begin{tabular}[c]{@{}l@{}} (\faArrowUp) Ability to dig-deeper into requirements \\ (\faArrowUp) Ability to resolve complicated requirements \\ (\faArrowUp) Increase the team engagement \\ (\faArrowUp) Opportunity to learn new things (e.g. technology, domain) \end{tabular} & \begin{tabular}[c]{@{}l@{}}  Excitement-\\Seeking (E5)\end{tabular} & Extraversion \\\\

\begin{tabular}[c]{@{}l@{}} Coping with \\others \end{tabular} & \begin{tabular}[c]{@{}l@{}} (\faArrowUp) Coping with other team members result in coming up with the \\best outcome\\ (\faArrowUp) Easy team management \\ (\faArrowUp) Easy to handle customers \\ (\faArrowUp) Easier to connect with the customers to get the requirements and \\synchronise them with various user levels \end{tabular} & \begin{tabular}[c]{@{}l@{}} Compliance/\\Cooperation\\ (A4)\end{tabular} & Agreeableness \\\\

\begin{tabular}[c]{@{}l@{}} Trust\\-worthiness\end{tabular} & 
\begin{tabular}[c]{@{}l@{}} (\faArrowUp) Leads to deliver projects on time without delays \\ (\faArrowUp) Helps to resolve internal conflicts within the team \\ (\faArrowUp) Getting complete requirements due to the trust between customers\\ and the team  \end{tabular} & Trust (A1) & Agreeableness \\\\

\begin{tabular}[c]{@{}l@{}} Considerate\\ nature\end{tabular} & 
\begin{tabular}[c]{@{}l@{}} (\faArrowUp) Ability to get the customer involvement throughout the project \\ (\faArrowUp) Reduce conflicts between the team members \\ (\faArrowUp) Improve the working comfort within the team, result in improving the\\ progress of the team's work \\ (\faArrowUp) supportive team members, helpful in completing the tasks correctly\\ on time  \end{tabular} & Altruism (A3) & Agreeableness \\\\

\begin{tabular}[c]{@{}l@{}} Organized\\ nature\end{tabular} & \begin{tabular}[c]{@{}l@{}} (\faArrowUp) Helpful in completing projects with strict budget and timelines \\ (\faArrowUp) Impacts the quality of the deliverables because they lead the to\\ work with a plan \\ (\faArrowUp) Always prepared so that they can ask more related questions from\\ the customers  \end{tabular} & \begin{tabular}[c]{@{}l@{}}  Order/ Order\\-0liness (C2) \end{tabular} & \begin{tabular}[c]{@{}l@{}} Conscien\\-tiousness \end{tabular} \\\\

\begin{tabular}[c]{@{}l@{}} Responsible\\ nature\end{tabular} & \begin{tabular}[c]{@{}l@{}} (\faArrowUp) Successful project delivery as they tend to complete their tasks\\ well on time \\ (\faArrowUp) Helpful in critical releases as they tend to work with the team until \\the work is done \\ (\faArrowUp) Successful project implementation, increasing customer satisfaction \end{tabular} & \begin{tabular}[c]{@{}l@{}} Dutifulness\\ (C3) \end{tabular}& \begin{tabular}[c]{@{}l@{}} Conscien\\-tiousness \end{tabular} \\\\

\begin{tabular}[c]{@{}l@{}} Less \\impulsiveness \end{tabular} & \begin{tabular}[c]{@{}l@{}}  (\faArrowUp) Not jumping into conclusions at the initial state of requirements gathering \\ (\faArrowUp) Reduce rapid requirement changes \\ (\faArrowUp) Important in gathering full requirements, that ends up delivering \\complete project \\ (\faArrowUp) Less complications of budget and schedule due to less number \\of changes in the later phase of the project \end{tabular} &  \begin{tabular}[c]{@{}l@{}} Deliberation/\\ Cautiousness \\(C6)\end{tabular} &  \begin{tabular}[c]{@{}l@{}} Conscien\\-tiousness \end{tabular} \\\\

\begin{tabular}[c]{@{}l@{}} Wanting to \\do the best \end{tabular} & \begin{tabular}[c]{@{}l@{}} (\faArrowUp) Identify best possible ways to provide the solution (the perfect solution) \\ (\faArrowUp) Less number of errors/ missing requirements \\ (\faArrowUp) Provide the best solution to the customers, increasing their satisfaction  \end{tabular} & \begin{tabular}[c]{@{}l@{}}Competence/ \\Self-Efficacy\\ (C1) \end{tabular} &  \begin{tabular}[c]{@{}l@{}} Conscien\\-tiousness \end{tabular}\\\\

\begin{tabular}[c]{@{}l@{}} Willingness \\to discuss \end{tabular} & \begin{tabular}[c]{@{}l@{}} (\faArrowUp) Constant discussions with the team, helpful in maintaining the project \\scope with the time \\ (\faArrowUp) Helpful to carry forward the project into the same direction \\ (\faArrowUp) Reduce conflicts between team members as well as with the customers \\ (\faArrowUp) Helpful in requirements specification and prioritization \end{tabular} &  \begin{tabular}[c]{@{}l@{}} Ideas/\\ Intellect (O5) \end{tabular} & \begin{tabular}[c]{@{}l@{}} Openness to\\ Experience \end{tabular}\\\\

\begin{tabular}[c]{@{}l@{}} Willingness\\ to change \end{tabular} &
\begin{tabular}[c]{@{}l@{}} (\faArrowUp) Prefer to change lead to explore, do experiments and change the\\ approaches accordingly \\ (\faArrowUp) helpful in creating innovative ideas \\ (\faArrowUp) Lean new technologies and techniques, to create best solutions \\ \end{tabular} & \begin{tabular}[c]{@{}l@{}} Actions/\\Adventu\\-rousness (O4)  \end{tabular}& \begin{tabular}[c]{@{}l@{}} Openness to\\ Experience \end{tabular}\\\\

\begin{tabular}[c]{@{}l@{}} Open-\\mindedness \end{tabular} &
\begin{tabular}[c]{@{}l@{}} (\faArrowUp) Ability to think out of the box, resulting in creating unique solutions \\ (\faArrowUp) Ability to think beyond given requirements \\ (\faArrowUp) Getting alternative ways of completing difficult tasks/ requirements  \end{tabular} & \begin{tabular}[c]{@{}l@{}}  Fantasy/ \\Imagination\\ (O1)\end{tabular} & \begin{tabular}[c]{@{}l@{}} Openness to\\ Experience \end{tabular} \\\\

\begin{tabular}[c]{@{}l@{}} Moody \& \\ anxious \\behaviour\end{tabular} & 
\begin{tabular}[c]{@{}l@{}} (\faArrowDown) Tend to overthink about requirements from the initial phase, resulting in\\ incomplete requirements gathering \\ (\faArrowDown) Lead to create confusions between the team and the customers \\ (\faArrowDown) Increase conflicts within internal team members \end{tabular} & Anxiety (N1) & Neuroticism \\\\

\begin{tabular}[c]{@{}l@{}} Easily \\getting \\annoyed\end{tabular} & 
\begin{tabular}[c]{@{}l@{}} (\faArrowDown) Leads to lose important customers as they tend to interact with them \\in aggressive/rude manner\\ (\faArrowDown) Creates lots of conflicts between team members \\ (\faArrowDown) Leads to impact the overall performance of the team \end{tabular} & \begin{tabular}[c]{@{}l@{}}Angry Hostility\\/ Anger (N2) \end{tabular} & neuroticism \\\\

\begin{tabular}[c]{@{}l@{}} Insecure\\ feeling\end{tabular} & 
\begin{tabular}[c]{@{}l@{}} (\faArrowDown) Team members' insecure feeling hinders the opportunity to try new \\technologies in the projects, resulting in the overall outcome of the project \\ (\faArrowDown) Stick with their preferred development areas, result in not getting the\\ best solution \\ (\faArrowDown) Increase the team conflicts as they are not cooperating with others \end{tabular} & \begin{tabular}[c]{@{}l@{}} Vulnerability\\ (N6)\end{tabular} & Neuroticism \\
  \\ \bottomrule 
\end{tabular}}%
\end{table*}

\par \faArrowCircleUp /\faArrowCircleDown\hspace{0.1cm}\textbf{Friendly/ outgoing nature:} \textcolor{black}{Team members with a personality facet of having a friendly/ outgoing nature play an important role when dealing with customers.} They are the key people who initiate and conduct discussions with customers, making them comfortable with the team to provide actual, in-depth requirements. Having friendly team members is helpful in dealing with customers with constant requirement changes and providing unrealistic or impractical requirements where they can lead the team in such circumstances to reduce the conflicts between the team and the customers. Moreover, they are beneficial in keeping the team's energy and carrying forward the teamwork positively: 
\faCommentsO\emph{``I have been observing this particular team member of my team who is always friendly and has some energy around him. He works with customers in a very friendly manner. So we didn't have many conflicts with the customer because he handles them" - INT01 [Lead BA].}\\
\faCommentsO\emph{``The outgoing members in the team can really drive the rest of the team, keeping the energy of the team" - INT04 [Business Analyst].}\\
However, \textcolor{black}{it was also mentioned that a team member's friendly/ outgoing nature personality facet can sometimes create challenging situations when they are involved in RE activities.} This can reduce the customers' involvement in RE activities, where they might not get enough opportunity to express their opinions. This may lead to incomplete requirements, creating requirements gaps and conflicts between the customers and the team. 
\faCommentsO\emph{``when the client is not getting enough chance to speak because this person is very friendly and always talking. So it can also impact creating gaps between the team and the clients" - INT10 [Business Analyst].}

\par \faArrowCircleUp /\faArrowCircleDown\hspace{0.1cm}\textbf{Leading \& controlling nature:} Leading \& controlling team members has a mixed impact when involved in RE activities. %where sometimes their personality facet of a leading \& controlling nature impacts positively. 
For example, according to INT02, who is working as a business analyst \& project manager, mentioned that it is a really important aspect to manage the internal team and lead the whole team towards one goal: \faCommentsO \emph{``So when people are less working as a team, I find it is important to have that one person in the team who is working as a team leader, managing the team and taking them to one page" - INT02 [Business analyst/Project manager])}\\
However, \textcolor{black}{many participants pointed out that they have experienced various challenges of having team member(s) with a leading \& controlling personality facet,} such as losing interest in the project, internal team conflicts, and not getting the best outcome due to over-power others' opinions: (E.g. \faCommentsO\emph{``we are working on one particular project, and we were doing equal contribution. But, he tried to stamp out my opinions. So, you tend to lose interest" - INT08 [Software engineer]}). \\
These characteristics relate to the \textbf{extraversion} dimension, where the `friendly/ outgoing' nature can be mapped onto the friendliness facet, whereas the `leading \& controlling' nature can be mapped into the assertiveness facet. 

\par \faArrowCircleUp \hspace{0.1cm}\textbf{Coping with others:} \textcolor{black}{Team members' ability to cope with others' personalities has a favorable impact when they participate in RE activities. }It specifically impacts the teamwork and overall performance of the team, as when team members are willing to cope with others to work, they come up with the best outcome. Moreover, managing the team and handling the customers is easier when the team is willing to cope with others:  
\faCommentsO\emph{``The members should be able to cope with others, you know, that improve our work, and we were able to give what client exactly expected" - INT11 [Lead BA]}.

\par \faArrowCircleUp \hspace{0.1cm}\textbf{Trustworthiness:} \textcolor{black}{Trust is needed within the team and with the customers, and this personality facet impacts the team significantly when involved in RE activities. }The trustworthiness of the team members leads to delivering the projects on time without delays. It helps to resolve internal conflicts within the team for the leads and the management. Trust between the team and the customers leads to getting complete requirements as customers tend to be more open when they trust the team: 
\faCommentsO\emph{``some kind of trust as well, we trust that they will deliver their tasks on time so that we can complete our project on time. Actually, that happened in most of my projects" - INT07 [Software engineer]}.\\
Both ``coping with others" and ``trustworthiness" characteristics can be mapped with the \textbf{agreeableness} dimension, where these two characteristics can be linked with cooperation and trust facets, respectively.

\par \faArrowCircleUp\hspace{0.1cm}\textbf{Organized nature:} Organized team members are very important when they are involved in RE activities. It helps to complete projects with strict budgets and timelines as the team is well prepared and organized, impacting the quality of the deliverables. A personality facet of an organized nature leads the team to work with a plan and is always prepared:  
\faCommentsO\emph{``you need to be organized so that your team has a plan and can work for it. In that project, we were able to complete it on time" -INT09 [Lead human-centred designer]}. 

\par \faArrowCircleUp\hspace{0.1cm}\textbf{Responsible nature:} Responsible team members are also a key to successful project delivery as they tend to complete their tasks well on time. Team members with a responsible nature personality facet are important when there are critical releases, as they tend to work with the team until the work is done. For example, INT07, a software engineer, mentioned that he had experienced the consequence of having an irresponsible team member who was always postponing the schedule, resulting in poor project implementation and customer dissatisfaction: 
\faCommentsO\emph{``We suggested some features, and he was like, he will do later, and he is postponing the schedule, but never implementing a good project, and the customer was not happy at the end" - INT07 [Software engineer]}.\\
These characteristics can be mapped onto the \textbf{conscientiousness} dimension, where the ``organized nature" can be mapped to the orderliness facet and the ``responsible nature" can be mapped to the dutifulness facet.

\par \faArrowCircleUp\hspace{0.1cm}\textbf{Willingness to discuss:} \textcolor{black}{Team members who have a willingness to discuss personality facet are helpful in completing RE activities and the overall project successfully.} Their willingness to have discussions within the team helps to maintain the project scope with time. For example, when all the members do not appear to be in the same direction, it would be great to have team discussions constantly to drive the team within the project scope: 
\faCommentsO\emph{``as a team, moments where we don't appear to be all are going in the same direction, we will have conversations often and lead the team to complete what is within the scope" - INT09 [Lead human-centred designer]}.

\par \faArrowCircleUp\hspace{0.1cm}\textbf{Willingness to change:} \textcolor{black}{Team members' willingness to change personality facet is also highly advantageous when they are involved in RE activities.} This leads them to explore more, do experiments and change their approaches accordingly. For example, when there is a technology change (upgrade) within the team, these team members accept the change as they are willing to explore and learn new things:  
\faCommentsO\emph{`` if you are actually doing requirement engineering, you need to have the ability to change, you should know how to go forward with new technologies, techniques and you have to have that desire to learn and cope with new things" - INT05 [Software engineer \& application consultant]}.\\
The characteristics of ``willingness to discuss" and ``willingness to change" can be mapped onto the \textbf{openness to experience} dimension, which can be related to the ideas/ intellect facet and actions/ adventurousness facet, respectively. 

\par  \faArrowCircleDown\hspace{0.1cm}\textbf{Moody \& anxious behaviour:} \textcolor{black}{ 7 participants mentioned that moody \& anxious personality facet of the team members may create challenging situations when they are involved in RE activities.} These team members tend to overthink requirements (e.g. how will we implement this?) in the first phase, not focusing on gathering complete requirements.
They may also create confusion between the team and the customers and conflicts among internal team members: 
\faCommentsO\emph{``if a person is moody or anxious, that may negatively impact carrying out requirement engineering activities, like they will be anxious and think about implementation work all the time so that they may miss important requirements" - INT02 [Business analyst \& Project manager]}.

\par \faArrowCircleDown\hspace{0.1cm}\textbf{Easily getting annoyed:} \textcolor{black}{The completion of RE activities and the overall projects have been hindered due to team members who are easily annoyed when they are involved in RE activities.} This personality facet leads them to lose valuable customers as they tend to interact with customers in a rude/aggressive manner. Also, it creates lots of conflicts among the team members. When questioned about their less cooperation with the team, they easily get annoyed, which will eventually impact the team's overall performance: 
\faCommentsO\emph{``there was a guy; he was very aggressive towards the problems and the people, he will get angry at customers. So the customers didn't want to work with him" - INT05 [Software engineer \& application consultant]}\\
Both characteristics can be related to the \textbf{neuroticism} dimension where ``moody \& anxious behaviour" can be mapped with anxiety and ``easily getting annoyed" can be mapped with anger facets, respectively.\\ 

\begin{itemize}
    \item \textbf{Impact related to clients/customers/end users' personality}
\end{itemize}

Twelve participants mentioned that clients/ customers/ end users' personalities can influence them when they are involved in RE activities, while the rest (3) mentioned that they do not believe that clients/ customers/end-users personalities have an impact when they are involved in RE activities. However, their insights are limited and quite similar to what they have expressed about the impact of their team members' personalities on RE activities. Some of the key findings related to the impacts of customers/clients/end users' personalities on RE activities are discussed as follows;

\par \faArrowCircleUp /\faArrowCircleDown\hspace{0.1cm}\textbf{Willingness to collaborate:} This can be mapped onto the cooperation facet in the agreeableness dimension, where customers' interest in working collaboratively with the team has a mixed impact on conducting RE activities. It helps to have a good discussion with the customers where practitioners can understand what they actually need and gather accurate, complete requirements. It reduces the number of requirements changes in the later phase of development as they are involved in the project from the beginning:  
\faCommentsO\emph{``It's very important in requirement engineering to have good discussions with the users; when they are working with us together, we can understand what they want, we can get all the requirements" - INT03 [IT development \& re-engineering lead]}.\\
However, \textcolor{black}{it was also mentioned that the customers' collaborative nature can sometimes be challenging to complete RE activities.} Due to their collaborative nature, they can often change the requirements, resulting in disruptions to the ongoing work and frustrating the software team. 
\faCommentsO\emph{``I have sometimes experienced their over-connection with the team disturbing our work. You know, they want to involve in everything, and then they keep changing and demanding, sometimes it is frustrating at the end of the day" - INT14 [Lead BA]}. 

\par \faArrowCircleUp\hspace{0.1cm}\textbf{Working with a plan:} \textcolor{black}{Customers who always work with a plan have a favourable impact on RE activities, thereby simplifying the tasks for software practitioners}. These customers have already thought of the requirements and know what they want, which leads the software practitioners to get clear requirements: 
\faCommentsO\emph{``So it's always great to work with clients who have planned what they want, they've thought it through, and we get the requirements easily" - INT10 [Business Analyst]}.\\
\faCommentsO\emph{``I worked with a foreign customer who is very organized, she has planned everything from the beginning, you know, she is like, we want these, these, very clear and it is easy for us. We can directly get the requirements and start the work" - INT15 [Senior software engineer]}.\\
This characteristic can be mapped with the self-discipline facet in the \textbf{conscientiousness} dimension.  

\par  \faArrowCircleUp\hspace{0.1cm}\textbf{Willingness to change:} \textcolor{black}{Customers who do not always stick to their beliefs and are open to change tend to have a beneficial impact on work when they are involved in RE activities. }Customers with this personality facet are helpful in changing requirements due to technical feasibility. They are open to considering suggestions from software teams, such as changing impractical requirements. This characteristic can be mapped with the actions/ adventurousness facet in the \textbf{openness to experience} dimension: 
\faCommentsO\emph{``There are some clients who are happy with what we suggest; they understand technical difficulties and accept changes. Then, we also try to give our best" - INT12 [Senior project manager]}.

\par  \faArrowCircleUp /\faArrowCircleDown\hspace{0.1cm}\textbf{Expressive nature:} \textcolor{black}{Customers with an expressive nature personality facet is greatly advantageous in carrying out of RE activities. It helps the team to get clearly defined requirements from the customers' end, and practitioners do not want to put extra effort into digging deep into the requirements to find out more about missing requirements. }They are not reluctant to say what they want and always work for it: 
\faCommentsO\emph{``if your customer is an expressive person, you really don't have to push them to give you information or detailed requirements. Because they would go out of their way to share it with you" - INT10 [Business analyst]}.\\
\textcolor{black}{However, it was also mentioned that this expressive nature personality facet can sometimes lead to challenging circumstances, such as creating project scope creep.} This characteristic can be related to the activity level facet in the \textbf{extraversion} dimension:  
\faCommentsO\emph{``customer who knows what he wants, keep expressing, keep demanding that this needs to be done and followed by B, C, D, and that could lead to scope creep, and we have to deal with those situations" - INT10 [Business analyst]}.
\par \faArrowCircleDown \hspace{0.1cm}\textbf{Reserved nature:} \textcolor{black}{Having a reserved customer can usually be challenging for the team when they are involved in RE activities. }The software practitioners involved with customers with this personality facet need to put extra effort into gathering requirements, which will also be time-consuming. The reserved customers tend not to provide their opinions, leading to incomplete requirements and conflicts with the software team members. This can be mapped with the self-consciousness facet in the \textbf{neuroticism} dimension: 
\faCommentsO\emph{``users with high emotionally and reserved, usually, they only say about one aspect at one time, so lots of time has to be given" -INT03 [IT development \& re-engineering lead]}.\\
\faCommentsO\emph{``I think sometimes they didn't at least raise their voice and ask questions or give their opinion. We didn't know whether complete what they expected and there were lots of confusions" - INT06 [Technical team lead]}.

\par The interviewed practitioners who believe there was no impact of customers' personality on RE explained the reasons for their opinions. This included that they need to do their job and should not get affected by the behaviours of external parties:  \\
\faCommentsO\emph{``they are external parties, and we cannot get affected by their behaviours" - INT05 [Software engineer \& application consult]}.\\
\faCommentsO\emph{``it's nothing major, you know, related to clients, no major complex, I think it is a minor thing, and our job is to give what they want" - INT06 [Technical team lead]}. \\
However, they consider their personality and team members' personalities to be important, and these do impact RE activities compared to their customers' personalities.

\subsection{Strategies to overcome some of the challenges of diverse personalities} \label{4.4}

\textcolor{black}{Some practitioners shared the strategies that they use to overcome some of the challenges occurred due to diverse personalities when involved in RE activities. These strategies (S) are used from the initial phase of the requirement engineering to the later phase of the projects and are mainly used within their teams. However, the purpose of these strategies is not to try and fix personalities of the software teams, but rather to provide proper understanding of the impact of personality on RE activities. }
\textcolor{black}{Not all of the strategies may be useful, or even applicable, in different contexts and for different teams, but sharing them here collectively will help practitioners understand what might work for them in their contexts. Further, when applying these strategies, we recommend consulting each team member, taking into account their preference with due care, as individual personality is a sensitive matter.
A summary of key strategies they reported are as follows}:

\par \textbf{S1: Changing the task allocations:} \textcolor{black}{The majority of the interviewed practitioners (10 out of 12) mentioned changing the task allocations of the team members as their initial strategy to overcome some of the challenges they faced due to diverse personalities when involved in RE activities. This can be done by resetting team responsibilities and assigning tasks that fit their personality. For example, INT04, a business analyst, mentioned that sometimes resetting the whole team's responsibilities and bringing the team back to a different state is important to address the challenging situations arising from the team's diverse personalities.} INT09, a lead human-centred designer, mentioned that assigning tasks that fit team members' personalities would be beneficial in gathering requirements without conflicts: \faCommentsO\emph{``I think because a lot of my work tends to involve interviewing people to gather requirements, some of the more introverted team members are sometimes shy about talking to other people, especially since they're not confident in dealing with senior stakeholders and gathering their requirements. And sometimes, they're not confident in speaking to users. So in that degree, a more extrovert personality is good for those types of tasks" - INT09}.\\
Creating a  working environment to reduce conflicts by applying different approaches to deal with differently opinionated people internally and externally also helps to overcome some of these challenges. However, it was highlighted that changes to the task allocations should be done with proper consideration: \faCommentsO \emph{``We ended up giving her a different role...she had done something similar in her previous role. So it was probably something where she felt a little bit more comfortable"  -INT12 [Senior project manager]}.

\par \textbf{S2: Conducting detailed discussions:} \textcolor{black}{Some of the participants (INT04, INT07, INT09, INT12 \& INT15) mentioned that conducting in-depth, cordial conversations with team members assist in overcoming some of the challenges arising from diverse personalities during RE activities.} It helps identify the disconnections within the team, understand each team member's problems and bring everyone on the same page, resulting in solving team conflicts and making them complete the tasks within the scope: \faCommentsO\emph{``my initial strategy is, identifying that there was a disconnection between the team and trying to just sit everyone down around the table and go……as a team, moments where we don't appear to be all going in the same direction, we will have conversations often, and lead the team to complete what is within the scope" - INT09 [Lead human-centred designer]}.

\par \textbf{S3: Right utilisation of diverse personalities:} \textcolor{black}{Ten out of twelve participants also mentioned that the 'right' utilization of diverse personalities within the team is one of the strategies employed by practitioners to overcome the adverse effects of personality on RE activities. }
They highlighted the importance of having a variety of personalities in their team. If they do not, it can also have a negative impact: \faCommentsO\emph{``if everyone within the team is an extrovert, you just end up with people talking over each other, and it is not good for the project either" - INT09 [Lead human-centred designer]}
%Since there are no 100\% positive or negative impacted personality characteristics, it is important to identify and utilize the right person for the activities: \emph{"But myself, I think I'm not so outgoing. But some of my team members are very outgoing, so we should use that to build a good relationship with customers. I also think you can work with people who may be nervous or insecure, as long as you give them the right support" - INT03 [IT development \& re-engineering lead]}.

\par \textbf{S4: Disseminate responsibilities:} \textcolor{black}{A few senior practitioners (INT01, INT12 \& INT15) shared that disseminating responsibilities and making the team members accountable for allocated tasks is an alternative approach employed by practitioners to address some of the personality-related challenges occurred when involved in RE activities.} 
By doing this, they try to give the team members a sense of accountability, so they tend to work together to complete the allocated tasks successfully. It improves team collaboration and, eventually, team performance as well: \faCommentsO\emph{``When we have meetings where we agree on what to do, I take action items, hold people to account, give them the responsibility and then they will work together to do things correctly"  - INT12 [Senior project manager]}.

\par \textbf{S5: Increase collaborations:} Some participants (INT01, INT07, INT09, INT12, \& INT15) highlighted that increasing collaboration is another strategy that can be used to alleviate the adverse effects of personality on RE. Here, they tend to increase the collaboration among team members by inculcating mutual trust and assigning tasks together to find a solution: \faCommentsO\emph{``try and instil a higher level of mutual trust and collaboration. That's always my first strategy, and most of the time that works" - INT09 [Lead human-centred designer]}.

\par \textbf{S6: Asking for management/higher-level support:} \textcolor{black}{Seeking management or higher-level support when necessary is another strategy practitioners use to counterbalance the influence of personality on RE activities (INT01, INT02, INT04, INT08 \& INT13). Especially in situations such as dealing with anxious/ insecure team members or dealing with tough customers, seeking management or higher-level support would be helpful in carrying out RE activities and the overall project.} However, it was also highlighted that developers should be able to identify and ask for support at the correct time before it is too late to be resolved: \faCommentsO\emph{`` I made an inquiry to my project manager and also my immediate supervisor about this concern because that particular resource was very anxious and heavily thinking, that made the customers kind of guilty and started not to reveal the requirements actually"  - INT01 [Lead BA]}.

\par \textbf{S7: Removing certain personalities in rare cases:} \textcolor{black}{A few participants (INT01, INT09 \& INT11) mentioned that treating according to the situation is one of their strategies used to overcome some of the challenges they faced due to diverse personalities when involved in RE activities.} This can vary from handling issues smoothly within the team and taking the necessary breaks to remove a person from the team. However, the ability to identify the correct situation is a must, and practitioners should take action accordingly: \faCommentsO\emph{``it had been a commercial situation, or would you just remove that person from the project, which is essentially what I did in that situation.... But my strategy is always to try and solve it in the first instance, where I try to solve it with the team smoothly. If the situation is not under control, then I have to act accordingly, and it's very rare that it happens" - INT09 [Lead human-centred designer].}

\begin{figure*}[t]
 \centering
  \includegraphics[width=\linewidth]{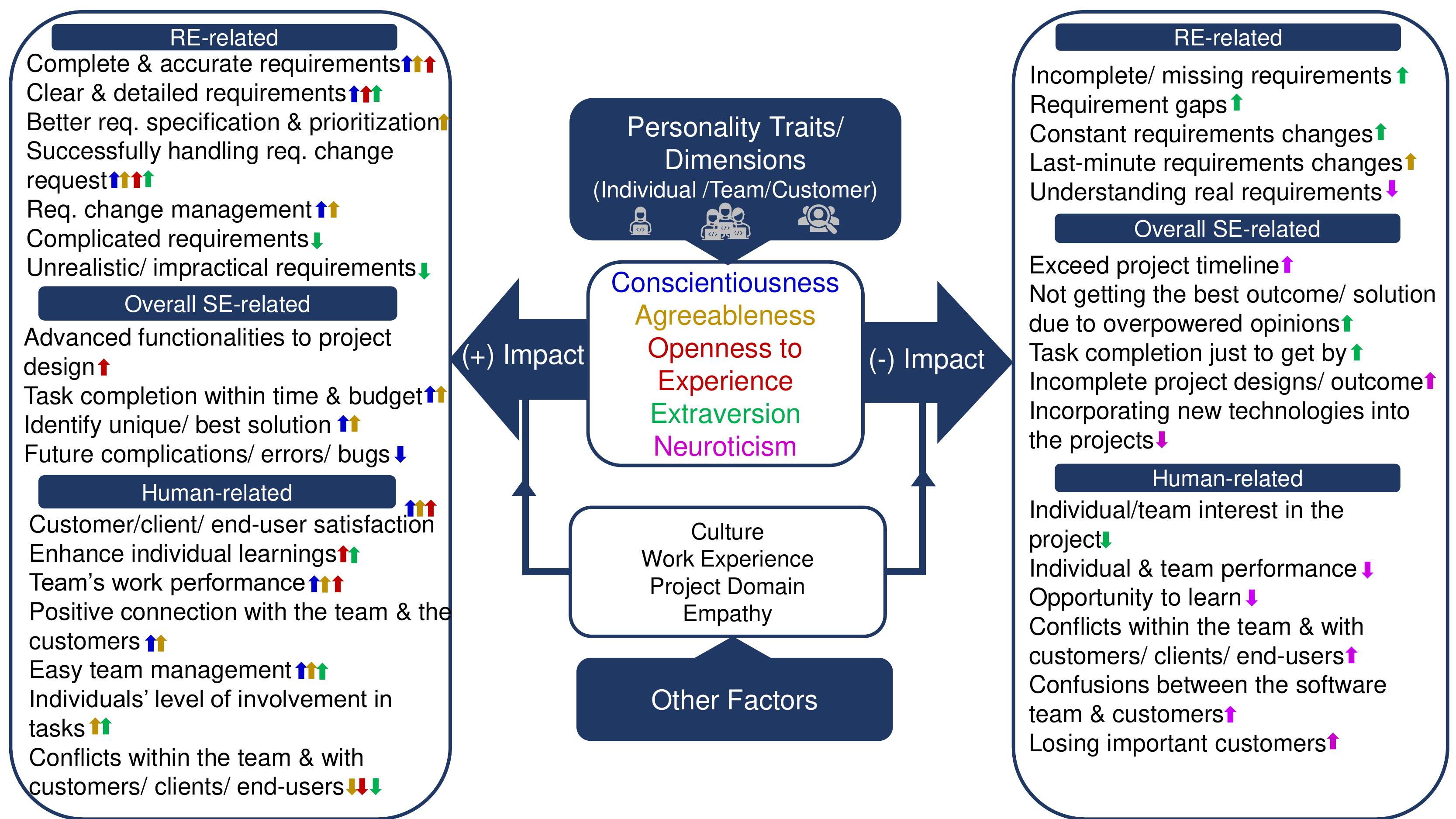}
  \caption{Overview of the impact of personality on RE/SE activities: The direction of the arrows indicates whether the impact is increasing  (\includegraphics[height=1em]{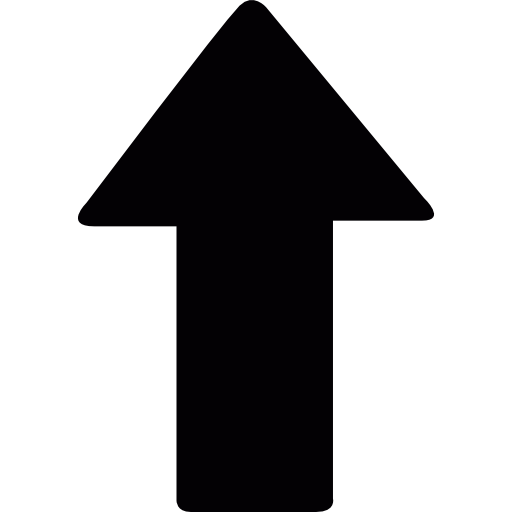}) or decreasing (\includegraphics[height=1em]{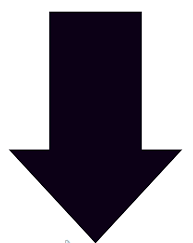}) and the colour of each arrow indicates the respective personality trait that causes the impact }
  \label{figure 06}
\end{figure*}

\section{Discussion} \label{section 5}
In this section, we discuss the key findings of our study, related to personality test-based survey and in-depth interviews, along with an overview of the impact of personality in RE activities based on the roles involved, such as software practitioners, their team members or customers/clients/ end users.

\subsection{Key findings}
Firstly, by analyzing the 50 personality profiles and data from 15 interviews, we have identified that the personality characteristics of the people involved in RE activities impact the successful completion of RE activities. \textcolor{black}{By analysing personality profiles, our aim was to gain preliminary insights into the personality traits of software practitioners who are predominantly involved in RE activities and identify any notable personality characteristics that may emerge from our 50 personality profiles. There, we have identified that \textbf{from our participants who are predominantly involved in RE activities, the majority have obtained high scores in agreeableness and conscientiousness characteristics}. This indicates that these practitioners have a strong interest in others' needs and well-being, are pleasant, sympathetic, cooperative, reliable, hardworking, set clear goals, and pursue them with determination. } \textbf{Extraversion and neuroticism characteristics are found at an average level (average-scored)}, indicating that some enjoy time with others while others prefer to spend time alone. This also means that stressful and frustrating situations are somewhat upsetting to them. However, they can generally get over them and cope with the situations. \textbf{An equal percentage of practitioners are found in either high or average levels related to openness to experience characteristics} indicating that 50\% of them enjoy novelty, variety, and change as they are curious, imaginative, and creative, and the other 50\% of the practitioners' thinking is neither simple nor complex. They prefer tradition but are also willing to try new things. 

\par Secondly, by analyzing our in-depth interviews, we have identified the impact of various personality characteristics related to software practitioners' personalities, their team members' personalities and customers' personalities. The majority of these characteristics can be mapped onto the five personality dimensions and related facets in the five-factor model (IPIP NEO-120). With that, we also identified how these personality characteristics are reported to impact RE activities and the overall software development process.  
As shown in figure \ref{figure 06}, we have summarized the key impacts of personality related to software practitioners', their team members and customers' personality characteristics. \textcolor{black}{These effects can be associated with the ways in which diverse personalities contribute to RE activities and/or the difficulties that arise as a result of diverse personalities. These have been classified into three distinct areas; namely \textbf{impacts related to requirements engineering (RE-related), impacts related to the overall software development process (overall SE-related) and impacts related to people involved in RE/SE (human-related).}} 
\par Although extraversion characteristics -- such as friendliness, outgoing and active nature -- are important in building a good relationship with customers, it is not the most important characteristic for RE activities (section \ref{4.3}). It was only considered as one such plus point, and other characteristics such as cooperativeness, open-mindedness, and hardworking nature are considered far more important (section \ref{4.3}). \faCommentsO\emph{``extraversion is not necessary; it's a plus trait; if you have it, it's good. But, if you don't, then it's perfectly fine; you still can do good in requirements gathering" - INT04 [Business analyst]}. Similarly, it was identified that the majority of neuroticism characteristics, such as anxiousness, insecure feelings, and anger, arise challenges that impact software practitioners and external stakeholders when involved in RE activities as well as the overall software development process. However, it was also mentioned that the right utilization of neuroticism characteristics of software practitioners could be beneficial as they tend to be deep-thinkers who might help the design phase. 

\par Apart from personality, we also identified several other factors that affect the identified impacts. The majority of our practitioners (11) mentioned  \textbf{individual's culture} as an important factor as they consider culture to have a strong connection with individuals' personalities.  The \textbf{work experience} of the practitioners, their \textbf{project domain}, and \textbf{empathy} have been mentioned as other factors that impact when involved in RE activities. Here, the work experience and project domain are interrelated as it was mentioned that more experience increases the exposure to areas/domains where individuals can handle various impacts of their previous experiences. Empathy was usually referred to when dealing with external people such as customers/ clients/ end-users when involved in RE activities. It was highlighted that empathizing with them makes the involvement in RE activities easier: (\faCommentsO\emph{``even to create the empathy because that has a huge factor I have seen, even myself in this learning curve, I practice being empathized with our clients, because they are dealing with other things, especially in the health sector and I think getting requirements are much more easier then"- INT13 [Lead BA]}). \textcolor{black}{From the interviews, we also identified several strategies that can be used to overcome some of the challenges arising from diverse personalities when engaging in RE/SE activities. Seven strategies were identified that could be applied to software teams, and the use of these strategies can vary with the project phase (section \ref{4.4}). For example, \emph{S4: disseminating responsibilities} can be used in the initial phases of the project to make the team members accountable for each task and reduce incomplete tasks at the end. The strategies identified were shared by the participants; these were the approaches they personally employed in their real-world projects to address some of the challenges arising from personality impacts. }

\subsection{Analysis of Interview Findings in light of Personality Profiles} \label{qualVquant}
\textcolor{black}{In our study, we collected 50 personality profiles of software practitioners who are mainly involved in RE activities, and of those, 15 software practitioners also participated in our follow-up interview study. We used the personality test-based survey to obtain preliminary insights into the personalities of the participants along with any notable traits/facets that may emerge and obtain an overview of their perspectives on the impact of personality on RE activities through an additional open-ended question at the end of the survey.} Using the standard IPIP-NEO 120 personality test, we obtained detailed information on the personality characteristics of participants via their personality profiles, and by analysing the open-ended answers, we identified that 94\% of them agreed that personality impacts RE activities and shared their perspectives and experiences about it. \\
\faCommentsO\emph{``the person's personality leads to a different approach to work and therefore to a different result in requirements engineering” - Senior Developer (P9 - survey participant)}. \\
The personality profiles provide detailed information about the participant’s own personality, which we mainly used to identify the most significant personality traits and facets among the participants (section \ref{4.2}). However, analysing these personality profiles and the open-ended answer does not provide in-depth insights into how personality impacts RE activities. Hence, along with the personality test, we invited participants to take part in semi-structured interviews with us. Of the 50 participants, 15 software practitioners volunteered to participate in the interviews. 

The interviews were focused on gaining deeper insights into the practitioner's perspectives of how personality characteristics impact RE activities, considering their own personality characteristics as well as their team members' and external stakeholders' (customers/clients/end users) personalities and \textcolor{black}{how they overcome some of the challenges arising from diverse personalities while working with their software teams.} The detailed analysis of their personality profiles shows that the majority (12 out of 15) of the participants have a high score in agreeableness, followed by conscientiousness and openness to experience traits (10 of 15 for each), and the majority of the participants have an average score in extraversion and neuroticism traits. The facets are being used to describe each personality trait in detail. The analysis of the facets shows that most of the facets in agreeableness (e.g. altruism facet, straightforwardness facet), conscientiousness (achievement-striving facet, competency facet), and openness to experience (e.g. ideas/intellect facet) personality traits have a high score compared to extraversion and neuroticism-related facets.  By analysing the interview data using STGT, we identified 16 personality-related characteristics that we linked to participant personalities (software practitioners’ individual personality – Table \ref{TABLE 4: Individual personality}). We mapped these characteristics identified via STGT onto personality facets and traits based on the definitions provided in the FFM of personality and the personality assessment test (IPIP-NEO 120) we used in our study.  There, we identified that most of the personality-related characteristics we discovered via STGT analysis that practitioners talked about referring to their own personality align with the findings of their personality profiles. For example, most of these identified characteristics (via STGT) are related to agreeableness (4 out of 16) and conscientiousness (4 out of 16), where these two traits have the most high-scored facets referring to their personality profiles. 
\par Apart from this confirmation, the interview findings provide an in-depth insight into the impacts of these personality characteristics when involved in RE activities, referring to their own personalities and considering others (team members and external stakeholders) involved in RE activities (section \ref{4.3}).  \textcolor{black}{From the interviews, it was also revealed that these impacts can lean in different directions, such as contributing to enhance the success or arising difficulties when doing RE activities, and how they overcome the difficulties through a set of strategies.}
 %Further, we identified that certain characteristics could impact positively or negatively based on the situations or the people involved in RE. 
 For example, ``considerate of others” has been identified as a personality characteristic as 14 out of 15 participants highlighted that their consideration towards others (e.g. team members) involved in RE activities helpful in solving conflicts and getting everyone’s opinions to get the best solution, which enhances the successful project outcome. This nature is directly related to the altruism facet in the agreeableness trait, and from the personality profile analysis, it is shown that 14 out of 15 participants have a high score in the altruism facet. For example, from their personality profile,  INT08 has a high score in the altruism facet. This was confirmed through their interview responses,
\faCommentsO\emph{``I am highly concerned about my team members; I personally talk to them to get their opinions and present them in the team discussions. It helps us to produce the best solution for our customers” -INT08 [Software engineer]}. 
\par Another important insight is that practitioners pointed out that the extraversion trait is only a plus point when involved in RE, and other personality characteristics, such as agreeableness and conscientiousness, are more important than extraversion characteristics (section \ref{4.3}). \\
\faCommentsO\emph{``extraversion is not necessary; it’s a plus trait; if you have it, it’s good. But, if you don’t, then it’s perfectly fine; you still can do good in requirements gathering” - INT04 [Business analyst]}.\\
\textcolor{black}{This, too, clearly shows from the personality profile analysis as the majority (10 out of 15) participants have an average score for the extraversion trait, and some of them even highlighted how extraversion characteristics become challenging when eliciting requirements referring to their team members.} \\
\faCommentsO\emph{``when the client is not getting enough chance to speak because this person is very friendly and always talking. So it can also impact creating gaps between the team and the clients” - INT10 [Business Analyst]}.

However, we also identified that some personality characteristics had not been brought up in the interviews, creating some differences between the results obtained from our STGT-based interview analysis and their personality profiles. For example, from the personality profile analysis, we identified that, in contrast, all the participants had a high score in the achievement-striving facet. However, only 6 out of 15 participants highlighted the importance of their hardworking nature and its impact on RE during interviews, where hardworking nature directly relates to the achievement-striving facet that is being used to describe the conscientiousness trait in FFM (e.g., ``I work hard” is one of the statements that are being used to measure the achievement-striving facet in IPIP-NEO 120 test). 

\textcolor{black}{Regarding the neuroticism characteristic, the majority of the participants discussed the challenges arising from neurotic characteristics such as moodiness, anxiety, insecurity, etc., and their impact on RE activities. However, only a few participants indicated that certain neuroticism-related characteristics, such as self-consciousness (without being bothered by difficult situations), positively impacted the design phase. For example, INT09 (Lead human-centred designer) stated that the right utilisation of neuroticism characteristics of software practitioners could be beneficial as they tend to be deep-thinkers who might help the design phase, highlighting the importance of maintaining a balanced mix of personalities within a software team.}

\textcolor{black}{Another observed difference was related to the neuroticism trait, which was seen to be a factor contributing to challenges during engagement in RE activities. For instance, we identified that 6 out of 15 participants have a high score in neuroticism in their personality profiles; however, only 4 out of 15 participants mentioned their anxious nature and how it creates difficulties when involved in RE activities, and only 3 out of 15 participants mentioned the challenge of involving in RE activities with their insecure feelings in their interviews. Also, we identified that when sharing their perspectives on the impact of their team members’ personalities, the participants talked more about the challenges faced due to neuroticism-related characteristics than they referred to themselves. This could be due to their personalities of having reluctance to explain how some of their own personality-related characteristics can be challenging in carrying out their work. Hence, there could be a potential bias as the interview findings are based on the participants' perceptions or diverse experiences they obtained when involved in RE activities.} 

\par When comparing our findings with existing literature, we identified that most of the studies incorporated the personality tests (IPIP-NEO 120/ IPIP-NEO 50)  in their surveys and used the personality test analysis to compare it with other factors such as decision-making style, team climate and etc. \cite{RN1602} \cite{RN3008}. However, from our mixed method approach, we identified that it is important to obtain deeper practitioners’ perspectives along with their personality profiles to better understand the impacts of personality in RE/SE activities and further investigations are required to identify ways to explain any differences observed between personality test results and interview results, as these differences may also happen due to personality differences. 

\subsection{Recommendations} \label{5.2}
Based on the findings from our study, we have identified several key challenges in the area of personality impacting RE, which need more focus. We have framed these as a set of recommendations, shared below, for software practitioners involved in RE activities and the wider SE research community for further research into the personality impact of RE/SE. 

\par \faHandORight\hspace{0.1cm}\textbf{Need for a balanced mix of personalities within software teams:} \textcolor{black}{According to our study findings, the majority of participants pointed out that having balanced personality characteristics in the team will help deliver the best output. Striking the right balance of personalities within a team is crucial for overall success in software development, including RE activities. For example, INT09 (Lead human-centred designer) stated that the right utilisation of neuroticism characteristics of software practitioners could be beneficial as they tend to be deep-thinkers who might help the design phase. As a result, our findings suggest that it's important for software practitioners and team leaders/project managers to emphasize the diversity of personalities within their teams. Meanwhile, SE/RE researchers can further explore identifying the most effective personality combinations for software development teams and devising strategies to overcome any potential downsides.}

%According to our study findings, there is no 100\% positive or negative impact of personality characteristics on RE activities.
%The majority of participants mentioned that having balanced personality characteristics in the team will help deliver the best output. Having the right mix of personalities in a team would make the whole software development process successful, including RE activities. Hence, it is important to pay more attention to having a diversity of personalities within the teams by software practitioners and team leads/ project managers while the SE/RE researchers can conduct more research on identifying the best personality combinations for software development teams while identifying approaches to mitigate the negative impacts.

\par \faHandORight\hspace{0.1cm}\textbf{Focusing on the impact of personality along with culture:} We identified that along with personality characteristics, most software practitioners mentioned that it is important to consider the \textit{culture} of the people involved in RE activities as it impacts individual personality. For example, it was mentioned that the culture of a particular country has a major impact on people's personalities. Some prefer to work with guidelines, and others prefer to work without rules and guidelines. These characteristics depend on their culture. \textcolor{black}{Hence, our findings suggest that it is important to further investigate the relationship between personality and culture through multi-national research. It would be interesting to study individuals from various cultures/ mixed cultures and analyze their implications in a global context.}

\par \faHandORight\hspace{0.1cm}\textbf{Apply and further investigate the strategies:} \textcolor{black}{We also identified a set of strategies (S) used by software practitioners to overcome some of        the challenges faced when involved in RE activities. We have identified seven strategies (section \ref{4.4}) that are used during RE-related activities, and these strategies are mainly for software team members. Hence, based on our findings, we recommend carrying out more research studies identifying and investigating these strategies that can be used for other software development activities and various contexts for the benefit of industry practitioners with due care as individual personality is a sensitive matter.}

\par \faHandORight\hspace{0.1cm}\textbf{Focusing on the impact of personality on individual and team performance:} \textcolor{black}{According to our practitioners, one of the major impacts of personality is their individual and team performance. They have mentioned that different personalities impact their overall work performance, and the impact can be related to how it enhances RE activities or challenges faced due to diverse personality characteristics. Hence, we suggest further investigating the impact of personality on individual and team performance, focusing on RE activities to identify the relationship between personality and performance when involved in RE activities.  }

\par \faHandORight\hspace{0.1cm}\textbf{Consideration of the impact of other factors when involved in RE activities:} While investigating the personality impact on RE activities, we identified a few other factors that could impact RE activities along with personality. As mentioned in recommendation 2, culture is the most mentioned aspect. Apart from that, practitioners have mentioned other human aspects such as empathy, geographic distribution, gender, communication skills, and other factors such as project domain and work experience. A very limited number of practitioners pointed out the impact of these aspects (one or two practitioners), and some of the practitioners believe that there is no impact of gender and age on RE activities: \faCommentsO\emph{``I've worked with people of different genders and ages. I don't believe that has a big impact" -INT08 [Software engineer]}. However, based on our findings, we recommend investigating these aspects in further empirical studies to get a more detailed understanding of the impact of these aspects on RE activities.

\par \faHandORight\hspace{0.1cm}\textbf{Performing empirical case studies in different software organizations:} \textcolor{black}{Since our findings are mainly based on our in-depth interview study, we suggest conducting more empirical case studies, revealing personality characteristics of individuals in various software teams in various software organizations globally.  This would enhance the understanding of how industry practitioners and organizations manage these effects and formulate improved strategies to enhance and counterbalance these effects. }

\section{Limitations and Threats to validity} \label{section 6}
 \textbf{Threats to external validity: } Our data collection does not possess an equal distribution of participants worldwide and does not represent the entire community of software practitioners involved in RE activities. \textcolor{black}{The generalizability of the results obtained by analysing the personality profiles is subject to the limitation of external validity due to 
 %relatively 
 small sample size (50 personality profiles). This study can be replicated with a larger sample size to validate or challenge our findings.} Further, the interviews were also conducted with 15 participants. However, each interview was between 50 -60 minutes long, providing rich insights into their perspectives, experiences and issues related to the impact of personality on RE activities. The majority of interview participants are from Sri Lanka, Australia and India. The details of all the participants and their organizations have been kept confidential as per the human ethics guidelines followed in this study. \textcolor{black}{The findings of our study are shaped by the participants' perspectives (those who took part in our study) and limited to the contexts represented by them such as organizations and their country of residence, limiting the generalization to the entire global SE community.} However, in practice, such generalization is unlikely achievable \cite{9513607}.

\par Regarding participants' job roles, there were 14 job roles/titles, and the interpretations of these roles can differ based on the organization. \textcolor{black}{However, the majority (40\%) of them were software engineers, followed by IT project managers (12\%) and lead business analysts (10\%).} In the personality test-based survey, we used a set of key RE-related job responsibilities to rate the practitioners' involvement (5-point Likert scale from ``Never" to ``Always") to see their involvement in RE activities. The majority of the participants were always or very often involved in given key-related job responsibilities. Interview participants described their responsibilities in detail, indicating that we collected data from our target participants. Since our main focus was to gather personality profiles using the personality test-based survey, \textcolor{black}{a set of demographic questions were tailored to elicit basic demographic information such as age, gender, country of residence, years of experience in the SE industry along with a set of questions to identify their involvement in RE activities. This helped us to collect data from our target participant group when collecting personality profiles and conducting interviews.}

%we included limited demographic questions with the standard personality test with all the instructions to complete the personality test.

\par In the interview study, we mainly focused on asking in-depth questions about participants' involvement in RE activities and their perspectives and experiences on the impact of personality when involved in RE activities. However, we have not framed the questions related to five personality traits and focused on asking their overall opinion on the personality impact on RE and then provided definitions/ examples as required. The participants have various experiences related to various domains, and their involvement in RE may vary with the context. Hence, we suggest conducting more empirical case studies focusing on different software organizations to investigate its impact on RE activities along with personality (section \ref{5.2} - recommendation 6).

\par  \textbf{Threats to internal validity: } \textcolor{black}{Although there are various personality assessment theories that have been used in the SE domain, from the comparative study \cite{RN2999}, FFM was considered the dominant, most suitable one referring to its validity and reliability. Further, as explained in section \ref{section 2.1}, among various personality tests that have been developed based on FFM, we used the IPIP-NEO 120 test in our study due to its acceptable reliability and practicality compared to other IPIP-NEO personality tests.} Following the STGT for qualitative data analysis, we have generated concepts and categories based on the codes. \textcolor{black}{After the first author conducted the initial coding and analysis, it was shared with all the other authors to discuss and resolve any different opinions.} Hence, all these codes, concepts and categories were collaboratively discussed and finalized by all the authors to try and overcome any potential biases. 

\textcolor{black}{Using payment for data collection of the personality test-based survey can also be a threat to the internal validity of the research as there may be fake responses when using participant recruitment platforms.} However, we have decided to use the Prolific tool after careful consideration following \cite{PALAN201822} and to avoid collecting fake/ incomplete responses, we only approved the payments for the participants after examining their responses to check whether they belong to our target participant group and if only they responded to each question (e.g., selected the ones with proper open-ended responses). \textcolor{black}{Further, to avoid potential bias of conducting the personality test-based survey prior to the interviews, we refrained from analyzing the personality profiles before the interviews took place. Additionally, for those interview participants who requested their personality profiles, we provided these profiles after the interviews were completed. This approach ensured that there were no preconceived notions or assessments of the interviewees' personalities prior to the interviews, thus minimizing the potential for bias.}

\par \textcolor{black}{We have provided all the standard instructions for the personality test and explained the personality-related terminologies during interviews with the participants. However, there could be still a potential bias on rating the personality test statements based on their self-perceptions. Further, the participants' understanding of personality can vary and based on that, the experiences they share during the interview may differ. The impacts of personality that we have identified may depend on their understanding of personality, their own perceptions and the strategies we have proposed to overcome some of the challenges faced due to diverse personalities may depend on how they are involved in such situations. These may also depend on particular contexts that they are involved in or their organizations.} Hence, by conducting more empirical case studies with software practitioners, more data can be collected related to organizations and contexts where software practitioners are involved in RE activities. This can eventually improve our findings.

\par Further, the personality profiles provide detailed information about our participants' overall personalities, and all the personality characteristics may not impact on how they carry out RE activities. For example, in the interviews, 06 out of 15 participants highlighted that their hard-working nature impacts RE activities and from the personality profile analysis, we identified that, in contrast, all the participants had a high score in the achievement-striving facet where hard-working nature directly relates to. This may create differences between their personality profiles and the perspectives on the impact of personality in RE, and we suggest further investigations to identify ways to explain these differences observed between personality test results and their perspectives.

\section{Conclusion} \label{section 7}
The findings of our study contribute to identifying the impact of personality of software practitioners, their team members and their external stakeholders (customers/clients/end-users) on RE activities. By analyzing personality profiles, we have identified that the majority of participants are high-scored in conscientiousness and agreeable nature and average in extraversion and neuroticism nature. The scores of the facets related to each personality dimension vary, and almost all the facets related to agreeableness and conscientiousness obtained high scores among participants. In the extraversion dimension, only cheerfulness and assertiveness are considerably high-scored, and all the other facets are average among participants, whereas, in neuroticism, almost all the facets have an average score among participants. This shows that the majority of the practitioners are highly determined, hard-working, cooperative, sympathetic individuals who have a strong interest in others' well-being. It also indicates that the majority of them are neither highly extraverted nor introverted individuals and tend to be that they enjoy time with others, but also time alone. An equal percentage of high and average-scored participants were identified as related to openness to experience nature. Regarding the openness to experience facets, the ideas/intellect, feelings, aesthetics and values/liberalism are high-scored facets compared to imagination and adventurousness facets, indicating that there is a balanced mix of high and average personality characteristics related to openness to experience traits among the practitioners. 
\par Through in-depth interviews, we have identified various personality impacts on RE activities due to software practitioners' individual personalities, their team members or customers/clients'/end-users personalities. \textcolor{black}{These impacts are mainly found with varying implications, which could either enhance the outcomes of RE activities or give rise to challenges when engaged in RE. and categorized as the impact on the RE activities (RE-related), impact on the overall software development (overall SE-related) or impact related to people involved in RE/SE (human-related). From these impacts, it shows that the majority of personality characteristics such as conscientiousness, agreeableness, openness to experience and extraversion have a beneficial influence on these three aspects, while certain neuroticism characteristics have a challenging impact on these aspects. We have also identified a set of strategies that can be applied to overcome some of the challenges faced due to diverse personalities when involved in RE activities }and the importance of focusing on other factors such as culture, work experience, project domain and empathy along with personality. The findings of this study will be beneficial for understanding the impact of personality on RE activities. \textcolor{black}{Whether agile or structured software development, having a balanced/ diverse software team is important to complete RE/SE-related activities successfully. However, observing various software teams in the industry would be interesting to see how they handle these situations in real-world scenarios to make the project successful. }

\begin{acknowledgements}
Hidellaarachchi was supported by Monash Faculty of IT PhD scholarships. Grundy and Hidellaarachchi are supported by ARC Laureate Fellowship FL190100035, and this work is also partially supported by ARC Discovery Project DP200100020.
\end{acknowledgements}
%If you'd like to thank anyone, place your comments here
%and remove the percent signs.
%\end{acknowledgements}

% Authors must disclose all relationships or interests that 
% could have direct or potential influence or impart bias on 
% the work: 
%
\section*{Declarations}
\textbf{Conflict of Interests:} The authors declare that they have no conflict of interest.\\
% \\
\textbf{Data Availability Statement:} All data generated or analysed during this study are included in this published article (and its supplementary information files).

% BibTeX users please use one of
%\bibliographystyle{spbasic}      % basic style, author-year citations
\bibliographystyle{spmpsci}      % mathematics and physical sciences
%\bibliographystyle{spphys}       % APS-like style for physics
%\bibliography{}   % name your BibTeX data base
\bibliography{output.bib}
% Non-BibTeX users please use
%\begin{thebibliography}{}
%
% and use \bibitem to create references. Consult the Instructions
% for authors for reference list style.
%
%\bibitem{RefJ}
% Format for Journal Reference
%Author, Article title, Journal, Volume, page numbers (year)
% Format for books
%\bibitem{RefB}
%Author, Book title, page numbers. Publisher, place (year)
% etc
%\end{thebibliography}

\appendix
\section{Appendix : Personality test based survey} \label{A}
\begin{footnotesize}
\textbf{Section 01: Demographic Information}
\begin{enumerate}
  \item How old are you? 
  \item How would you describe your gender? (Male/ Female/ Prefer to self-describe as:/ Prefer not to answer)
  \item Country of your residence?
  \item What is your highest educational qualification?
  \item How many years have you been in the software industry?
  \item Is eliciting/ analysing. prioritizing/ modelling/ managing software requirements a part of your job? (Yes/ No
  \item How often do you elicit/ analyse/ prioritize/ manage software requirements as a part of your job?
  \begin{itemize}
      \item Almost everyday/ -Couple of times a week/ -Couple of times a month/ -Couple of times a year (or very rarely)
  \end{itemize}
  \item What is your current job title/ job role?
  \item Your current job responsibilities include: (please rate the following based on your involvement): close-ended question with the likert scale  from \emph{``Never"} to \emph{``Always"}.
  \begin{itemize}
      \item Collaborate with the stakeholders to elicit requirements
      \item Documenting software requirements specifications according to standard templates
      \item Lead requirements analysis and verification
      \item Participate in requirements prioritization
      \item Manage requirements throughout the project
  \end{itemize}
  \item Is there are any other job responsibilities that you involve in apart from the above list, please mention:
  \item What types of domains* are you currently working on? Select all that apply. (Domains* - the subject area in which your current project belongs to)
  \begin{itemize}
      \item Health/ -Education/ -Finance/ -Transports \& logistics
      \item Government services/ --Others (please specify);
  \end{itemize}
  \item What type of software development methods you have majorly involved in? (Please
choose option(s) from below)
  \begin{itemize}
      \item Traditional (Waterfall)/ -Agile (Please specify)/ -Other: (Please specify)
  \end{itemize}
\end{enumerate}8

\textbf{Section 02: Personality Assessment}
\begin{enumerate}
\item This section contains phrases describing people's behaviours. Please go through the
instructions listed above before you rate the phrases.
Please indicate to what extent each of the following statements applies to you.\\
(Rate with the likert scale from \emph{``very inaccurate"} to \emph{``very accurate"}) - list of 120 statements can be found in \textbf{table \ref{personality test statements}}. 

\begin{table*}[]
%\centering
\caption{ Personality test statements}
\label{personality test statements}
\resizebox{\textwidth}{!}{%
\begin{tabular}{@{}|l|l|l|l|@{}} 
 \hline
 \begin{tabular}[c]{@{}l@{}} 1. I worry about\\ things \end{tabular}  
 & \begin{tabular}[c]{@{}l@{}} 31. I fear for \\the worst \end{tabular}  
 & \begin{tabular}[c]{@{}l@{}} 61. I am afraid of\\ many things\end{tabular} 
 & \begin{tabular}[c]{@{}l@{}} 91. I get stressed\\ out easily\end{tabular}\\ 
 
 \begin{tabular}[c]{@{}l@{}}  2. I make friends \\easily \end{tabular} 
 & \begin{tabular}[c]{@{}l@{}} 32. I feel comfortable \\around people \end{tabular}
 & \begin{tabular}[c]{@{}l@{}} 62. I avoid contacts\\ with others \end{tabular} 
 & \begin{tabular}[c]{@{}l@{}} 92. I keep others\\at a distance \end{tabular}\\ 
 
  \begin{tabular}[c]{@{}l@{}} 3. I have a vivid \\imagination \end{tabular} 
  & \begin{tabular}[c]{@{}l@{}} 33. I enjoy wild\\ flights of fantasy \end{tabular} 
  & \begin{tabular}[c]{@{}l@{}} 63. I love to \\daydream \end{tabular}
  &\begin{tabular}[c]{@{}l@{}} 93. I like to get\\ lost in thought\end{tabular}\\ 
 
 \begin{tabular}[c]{@{}l@{}} 4. I trust others \end{tabular}
 &  \begin{tabular}[c]{@{}l@{}} 34. I believe that\\ others have good \\intentions \end{tabular} 
 & \begin{tabular}[c]{@{}l@{}} 64. I trust what \\people say\end{tabular}
 & \begin{tabular}[c]{@{}l@{}} 94. I distrust people\end{tabular}\\
 
 \begin{tabular}[c]{@{}l@{}}  5. I complete tasks \\successfully \end{tabular} 
 & \begin{tabular}[c]{@{}l@{}} 35. I excel in what\\ I do  \end{tabular}
 & \begin{tabular}[c]{@{}l@{}}65. I handle tasks\\ smoothly \end{tabular} 
 &  \begin{tabular}[c]{@{}l@{}} 95. I know how to\\ get things done \end{tabular}\\
  
  \begin{tabular}[c]{@{}l@{}}  6. I get angry \\easily\end{tabular} 
  & \begin{tabular}[c]{@{}l@{}} 36. I get irritated \\easily  \end{tabular} 
  & \begin{tabular}[c]{@{}l@{}} 66. I lose my \\temper \end{tabular}
  & \begin{tabular}[c]{@{}l@{}} 96. I am not easily\\ annoyed\end{tabular}\\
  
   \begin{tabular}[c]{@{}l@{}} 7. I love large \\parties\end{tabular} 
   & \begin{tabular}[c]{@{}l@{}} 37. I talk to a\\ lot of different \\people at parties \end{tabular} 
   & \begin{tabular}[c]{@{}l@{}} 67. I prefer to be alone \end{tabular}
   & \begin{tabular}[c]{@{}l@{}}97. I avoid crowds\end{tabular}\\
   
   \begin{tabular}[c]{@{}l@{}}8. I believe in \\the importance of art \end{tabular} 
   & \begin{tabular}[c]{@{}l@{}} 38. I see beauty in \\things that others \\might not notice\end{tabular}
   & \begin{tabular}[c]{@{}l@{}} 68. I do not like\\ poetry \end{tabular}
   & \begin{tabular}[c]{@{}l@{}} 98. I do not enjoy \\going to art \\museums\end{tabular}\\
   
   \begin{tabular}[c]{@{}l@{}} 9. I use others \\for my own ends\end{tabular}  
   & \begin{tabular}[c]{@{}l@{}} 39. I cheat to \\get ahead \end{tabular}
   & \begin{tabular}[c]{@{}l@{}} 69. I take advantage\\ of others\end{tabular}
   & \begin{tabular}[c]{@{}l@{}} 99. I obstruct others’\\ plans \end{tabular}\\
   
   \begin{tabular}[c]{@{}l@{}} 10. I like to \\tidy up \end{tabular} 
   & \begin{tabular}[c]{@{}l@{}} 40. I often forgot \\to put things back \\in their proper place \end{tabular}
   & \begin{tabular}[c]{@{}l@{}} 70. I leave a mess\\ in my room\end{tabular}
   & \begin{tabular}[c]{@{}l@{}} 100. I leave my \\belongings around\end{tabular}\\
   
   \begin{tabular}[c]{@{}l@{}} 11. I often feel \\blue\end{tabular}
   & \begin{tabular}[c]{@{}l@{}} 41. I dislike myself \end{tabular} 
   & \begin{tabular}[c]{@{}l@{}} 71. I am often \\down in the dumps \end{tabular}
   &\begin{tabular}[c]{@{}l@{}} 101. I feel comfortable\\ with myself \end{tabular}\\
   
   \begin{tabular}[c]{@{}l@{}} 12. I take charge\end{tabular} 
   & \begin{tabular}[c]{@{}l@{}} 42. I try to lead\\ others \end{tabular} 
   & \begin{tabular}[c]{@{}l@{}} 72. I take control \\of things \end{tabular}
   & \begin{tabular}[c]{@{}l@{}} 102. I wait for \\others to lead the \\way \end{tabular}\\
   
   \begin{tabular}[c]{@{}l@{}} 13. I experience my \\emotions intensely \end{tabular} & \begin{tabular}[c]{@{}l@{}} 43. I feel others’ \\emotions  \end{tabular}
   & \begin{tabular}[c]{@{}l@{}} 73. I rarely notice \\my emotional reactions \end{tabular}
   & \begin{tabular}[c]{@{}l@{}} 103. I don’t understand \\people who get \\emotional \end{tabular}\\
   
   \begin{tabular}[c]{@{}l@{}} 14. I love to help\\ others\end{tabular} 
   & \begin{tabular}[c]{@{}l@{}} 44. I am concerned\\ about others \end{tabular} 
   &\begin{tabular}[c]{@{}l@{}} 74. I am indifferent\\ to the feelings of \\others \end{tabular} 
   & \begin{tabular}[c]{@{}l@{}} 104. I take no time\\ for others\end{tabular}\\
   
   \begin{tabular}[c]{@{}l@{}} 15. I keep my promises \end{tabular} 
   & \begin{tabular}[c]{@{}l@{}} 45. I tell the truth \end{tabular}
   & \begin{tabular}[c]{@{}l@{}} 75. I break rules \end{tabular}
   & \begin{tabular}[c]{@{}l@{}} 105. I break my \\promises\end{tabular}\\
   
   \begin{tabular}[c]{@{}l@{}} 16. I find it \\difficult to approach \\others \end{tabular} 
   & \begin{tabular}[c]{@{}l@{}} 46. I am afraid \\to draw attention to \\myself \end{tabular}
   & \begin{tabular}[c]{@{}l@{}} 76. I only feel \\comfortable with \\friends \end{tabular}
   & \begin{tabular}[c]{@{}l@{}} 106. I am not bothered\\ by difficult social\\ situations\end{tabular}\\
   
   \begin{tabular}[c]{@{}l@{}} 17. I am always \\busy\end{tabular}
   & \begin{tabular}[c]{@{}l@{}} 47. I am always on\\ the go \end{tabular}
   & \begin{tabular}[c]{@{}l@{}} 77. I do a lot in\\ my spare time\end{tabular}
   & \begin{tabular}[c]{@{}l@{}} 107. I like to take\\ it easy\end{tabular}\\
   
  \begin{tabular}[c]{@{}l@{}} 18.  I prefer variety \\of routine\end{tabular} 
  & \begin{tabular}[c]{@{}l@{}}48. I prefer to stick\\ with the things \\that I know \end{tabular}
  & \begin{tabular}[c]{@{}l@{}} 78. I dislike changes  \end{tabular}
  & \begin{tabular}[c]{@{}l@{}} 108. I am attracted \\to conventional ways\end{tabular}\\
  
  \begin{tabular}[c]{@{}l@{}} 19. I love a good \\fight \end{tabular} 
  & \begin{tabular}[c]{@{}l@{}} 49. I yell at people \end{tabular}
  & \begin{tabular}[c]{@{}l@{}} 79. I insult people\end{tabular}
  & \begin{tabular}[c]{@{}l@{}}109. I get back at \\others \end{tabular}\\
  
  \begin{tabular}[c]{@{}l@{}} 20. I work hard\end{tabular} & \begin{tabular}[c]{@{}l@{}} 50. I do more than \\what’s expected of me  \end{tabular}
  & \begin{tabular}[c]{@{}l@{}} 80. I do just enough\\ work to get by \end{tabular}
  & \begin{tabular}[c]{@{}l@{}} 110. I put little time \\and effort into my \\work\end{tabular}\\
  
  \begin{tabular}[c]{@{}l@{}} 21. I go on binges \end{tabular} 
  & \begin{tabular}[c]{@{}l@{}} 51. I rarely overindulge \end{tabular}
  & \begin{tabular}[c]{@{}l@{}} 81. I easily resist\\ temptations \end{tabular}
  & \begin{tabular}[c]{@{}l@{}} 111.  I am able to \\control my cravings\end{tabular}\\
  
  \begin{tabular}[c]{@{}l@{}} 22. I love \\excitement\end{tabular} 
  & \begin{tabular}[c]{@{}l@{}} 52. I seek adventure \end{tabular}
  & \begin{tabular}[c]{@{}l@{}} 82. I enjoy being \\reckless  \end{tabular}
  & \begin{tabular}[c]{@{}l@{}} 112. I act wild \\and crazy \end{tabular}\\
  
  \begin{tabular}[c]{@{}l@{}}23. I love to read\\ challenging material\end{tabular} 
  & \begin{tabular}[c]{@{}l@{}} 53. I avoid \\philosophical discussions \end{tabular}
  & \begin{tabular}[c]{@{}l@{}} 83. I have difficulty\\ understanding abstract \\ideas \end{tabular}
  & \begin{tabular}[c]{@{}l@{}} 113. I am not interested \\in theoretical discussions\end{tabular}\\
  
  \begin{tabular}[c]{@{}l@{}} 24. I believe that\\ I am better than others\end{tabular} 
  & \begin{tabular}[c]{@{}l@{}} 54. I think highly \\of myself \end{tabular}
  & \begin{tabular}[c]{@{}l@{}} 84. I have high opinion \\of myself\end{tabular}
  & \begin{tabular}[c]{@{}l@{}} 114. I boast about \\my virtues \end{tabular}\\
  
  \begin{tabular}[c]{@{}l@{}} 25. I am always \\prepared \end{tabular} 
  & \begin{tabular}[c]{@{}l@{}} 55. I carry out my\\ plans\end{tabular}
  & \begin{tabular}[c]{@{}l@{}} 85. I waste my time \end{tabular}
  & \begin{tabular}[c]{@{}l@{}} 115. I have difficult \\starting tasks \end{tabular}\\
  
  \begin{tabular}[c]{@{}l@{}} 26. I panic easily\end{tabular} 
  & \begin{tabular}[c]{@{}l@{}} 56. I become \\overwhelmed by events \end{tabular}
  & \begin{tabular}[c]{@{}l@{}} 86. I feel that I’m \\unable to deal with \\things\end{tabular}
  & \begin{tabular}[c]{@{}l@{}} 116. I remain calm \\under pressure\end{tabular}\\
  
  \begin{tabular}[c]{@{}l@{}} 27. I radiate joy\end{tabular}  
  & \begin{tabular}[c]{@{}l@{}} 57. I have a lot \\of fun \end{tabular}
  & \begin{tabular}[c]{@{}l@{}} 87. I love life \end{tabular}
  & \begin{tabular}[c]{@{}l@{}} 117. I look at the \\bright side of life\end{tabular}\\
  
  \begin{tabular}[c]{@{}l@{}} 28. I tend to vote\\ for liberal political \\candidates\end{tabular}
  & \begin{tabular}[c]{@{}l@{}} 58. I believe that \\there is no absolute\\ right or wrong \end{tabular}
  & \begin{tabular}[c]{@{}l@{}} 88. I tend to vote\\ for conservative political\\ candidates \end{tabular}
  & \begin{tabular}[c]{@{}l@{}} 118. I believe that \\we should be tough\\ on crime \end{tabular}\\
  
  \begin{tabular}[c]{@{}l@{}} 29. I sympathize with \\the homeless\end{tabular} 
  & \begin{tabular}[c]{@{}l@{}} 59. I feel sympathy\\ for those who are \\worse off than myself \end{tabular}
  & \begin{tabular}[c]{@{}l@{}} 89. I am not interested\\ in other people’s \\problems \end{tabular}
  & \begin{tabular}[c]{@{}l@{}} 119. I try not to\\ think about the needy\end{tabular}\\
  
  \begin{tabular}[c]{@{}l@{}} 30. I jump into things\\ without thinking \end{tabular} 
  & \begin{tabular}[c]{@{}l@{}} 60. I make rash \\decisions \end{tabular}
  & \begin{tabular}[c]{@{}l@{}} 90. I rush into things \end{tabular}
  & \begin{tabular}[c]{@{}l@{}} 120. I act without \\thinking \end{tabular}\\
  
 \hline
\end{tabular}%
}
\end{table*}

\item Do you think personality of the people involved in requirements engineering activities has an impact on it's activities? (Yes/ No)
\item Please explain your above answer briefly:
\item Do you wish to receive your individual personality profile? (Yes/No)
\item If Yes, please enter your name and email address:
\end{enumerate}
\end{footnotesize}

\section{Appendix: Interview Schedule} \label{B}
\begin{footnotesize}
\setlength{\parindent}{4pt}
\textbf{Demographic Information} 
\begin{itemize}
    \item Can you briefly tell me about yourself? 
    \item Considering your experience; how many years of experience do you have in the software industry?
    \item How many years of experience you have in requirement engineering in these years?
    \item Can you think of a past or a current software engineering project where you were engaged in RE activities? 
    \begin{itemize}
        \item Can you please tell me very briefly about that project?
        \item Were you developing a product or service?
        \item What software development methods are/were you using?
        \item What is the size and the composition of the team?
        \item What is/was your role on the project?
        \item What are RE-related activities were being performed on that project? 
        \item What was your involvement in those activities?
    \end{itemize}
\end{itemize}

\textbf{Views on the influence of personality in RE-related activities} 
\par \textbf{Explanation:} Personality can be simply described as a set of individual differences including personal habits, skills, memories, behaviours and social relationships that can be affected by social and cultural development of individuals. There are various personality models that can be used to measure individual personality and the one you all have used is Five Factor Model- one of the most used models in this domain. There are five broad dimensions used in common language to describe human personality as follows.\\
\textbf{Openness to Experience:} Someone who is high on openness to experience tends to appear as imaginative, broad-minded and curious whereas those at the opposite end of this spectrum prefer for routine and favouring conservative values\\
\textbf{Conscientiousness:} People who are high in conscientiousness tend to be hardworking, organized, able to complete tasks thoroughly and on-time, and reliable.  On the other hand, low conscientiousness relates to traits such as being irresponsible, impulsive and disordered\\
\textbf{Extraversion:}A person is considered an extravert if he/she feels comfortable in a social relationship, friendly, assertive, active and outgoing\\
Agreeableness: Refers to traits such as cooperativeness, kindness, trust and warms whereas people who are low on agreeableness tend to be sceptical, selfish and hostile\\
\textbf{Neuroticism:} Refers to the state of emotional stability. Someone who is low on Neuroticism tends to appear calm, confident and secure, whereas a high neuroticism individual tends to be moody, anxious, nervous and insecure. \\
Talking about an individual’s personality;
\begin{itemize}
    \item Have you experienced a person’s personality having an effect or impact on how they approach requirements engineering?
    \begin{itemize}
        \item If yes, can you share your experience?
        \item If not, why do you think it does not have any impact?
    \end{itemize}
    \item Can you think of a time where your own personality (or someone's) positively influenced how they carried out RE activities?  
    \begin{itemize}
        \item Who else did it impact and how?
    \end{itemize}
    \item Can you think of a time where your own personality (or someone's) negatively influenced how they carried out RE activities?  
      \begin{itemize}
        \item Who else did it impact and how?
    \end{itemize}
    \item Have you experienced your team members' personality influencing how they carried out RE activities?
    \begin{itemize}
        \item If yes, can you share your experience?
        \item If not, why do you think it does not influence?
    \end{itemize}
    \item Have you experienced your customer’s/ client's personality impacting RE activities?
     \begin{itemize}
        \item If yes, can you share your experience?
        \item If not, why do you think it does not influence?
    \end{itemize}
    \item Have you experienced combinations of different personality types impacting RE activities? 
    \begin{itemize}
        \item If yes, can you share your experience?
    \end{itemize}
    \item How do you handle various personality differences within your team?
    \item In your experience, are there any other human aspects/ human-centric issues that have an impact on performing RE activities?
       \begin{itemize}
        \item If yes, what are they? 
        \item How do they impact?
    \end{itemize}
    \item Any final thoughts about the impact of personality on RE-related activities?
\end{itemize}
 \end{footnotesize}
 
 \section{Appendix: Sample Personality Profile} \label{C}
 \textbf{IPIP-NEO Narrative Report –Personality Profile}( \textbf{Participant INT11})


\section*{Supplementary material}

# IPIP -NEO Narrative Report – Personality Profile

**NOTE:** This report is sent to you upon your request, and it is analysed using the standard IPIP-NEO 120 Personality testing tool developed based on the standard Five-Factor Model (FFM) of Personality.

This report estimates the individual's level on each of the five broad personality domains of the Five-Factor Model. The description of each one of the five broad domains is followed by a more detailed description of personality according to the six sub-domains/facets that comprise each domain.

*A note on terminology* – Personality traits describe relative to other people, the frequency or intensity of a person's feelings, thoughts, or behaviours. Possession of a trait is therefore a matter of degree. We might describe two individuals as "extraverts", but still see one as more extraverted than the other. This report uses expressions such as "extravert" or "high in extraversion" to describe someone who is likely to be seen by others as relatively extraverted. This report is generated based on the analysis of your answers to the personality test and classifies you as low, average, or high in a trait according to whether your score is approximately in the lowest 30%, middle 40% or highest 30% of scores obtained by people of your sex and roughly your age. Your numerical scores are reported and graphs as percentile estimates.

Please keep in mind that "low", 'average" and "high" scores on a personality test are neither absolutely good nor bad. A particular level on any trait will probably be neutral or irrelevant for a great many activities, be helpful for accomplishing some things and detrimental for accomplishing other things. As with any personality inventory/tool, scores and descriptions can only approximate an individual's actual personality. High and low score descriptions are usually accurate, but average scores close to the low or high boundaries might misclassify you as only average. On each set of six sub-domain scales, it is somewhat uncommon but certainly possible to score high in some of the sub-domains and low in others. In such cases, more attention should be paid to the sub-domain scores than the broad domain score. Questions about the accuracy of your results are best resolved by showing your report to people who know you well.

Dr John A. Johnson, the founder of IPIP-NEO 120 personality inventory, wrote the description of the five domains and thirty subdomains. These descriptions are based on extensive reading of the scientific literature on personality measurement. Hence, we would like to acknowledge Dr John A. Johnson for finding and introducing this personality inventory to the public domain.

## Extraversion

Extraversion is marked by pronounced engagement with the external world. Extraverts enjoy being with people, are full of energy and often experience positive emotions. They tend to be enthusiastic, action-oriented individuals who like to say "Yes!" or "let's go!" to opportunities for excitement. In groups, they like to talk, assert themselves and draw attention to themselves.

Introverts lack the exuberance, energy, and activity levels of extroverts. They tend to be quiet, low-key, deliberate and disengaged from the social world. Their lack of social involvement should not be interpreted as shyness or depression. The introvert simply needs less stimulation than an extrovert and prefers to be alone. The independence and reserve of the introvert are sometimes mistaken as unfriendliness or arrogance. In reality, an introvert who scores high on the agreeableness dimension will not seek others out but will be quite pleasant when approached.

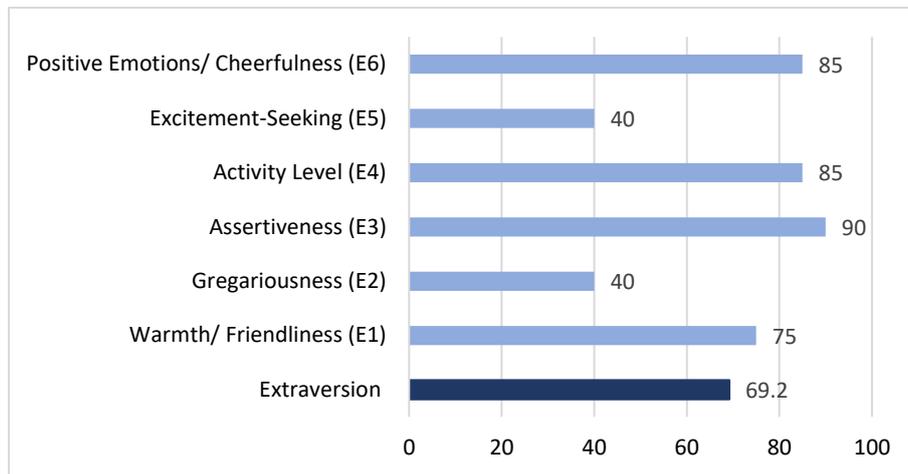

Your score on Extraversion is average, indicating you are neither a subdued loner nor a jovial chatterbox. You enjoy time with others, but also time alone.

*Extraversion Facets:*

- Friendliness (E1) – Friendly people genuinely like other people and openly demonstrate positive feelings toward others. They make friends quickly and it is easy for them to form close, intimate relationships. Low scorers on Friendliness are not necessarily cold and hostile, but they do not reach out to others and are perceived as distant and reserved. Your level of friendliness is high.
- Gregariousness (E2) – Gregarious people find the company of others pleasantly stimulating and rewarding. They enjoy the excitement of crowds. Low scorers tend to feel overwhelmed by, and therefore actively avoid large crowds. They do not necessarily dislike being with people sometimes, but their need for privacy and time to themselves is much greater than for individuals who score high on this scale. Your level of gregariousness is average.
- Assertiveness (E3) – High scorers' assertiveness like to speak out, take charge and direct the activities of others. They tend to be leaders in groups. Low scorers tend not to talk much and let others control the activities of groups. Your level of assertiveness is high.
- Activity Level (E4) – Active individuals lead fast-paced, but lives. They move about quickly, energetically, and vigorously and they are involved in many activities. People who score low on this scale follow a slower and more leisurely, relaxed pace. Your activity level is high.
- Excitement-Seeking (E5) – High scorers on this scale are easily bored without high levels of stimulation. They love bright lights and hustle and bustle. They are likely to take risks and seek thrills. Low scorers are overwhelmed by noise and commotion and are averse to thrill-seeking. Your level of excitement-seeking is average.
- Cheerfulness (E6) – This scale measures positive mood and feelings, not negative emotions (which are a part of the Neuroticism domain). Persons who score high on this scale typically experience a range of positive feelings, including happiness enthusiasm, optimism, and joy. Low scorers are not as prone to such energetic, high spirits. Your level of positive emotions is high.

**Agreeableness**

Agreeableness reflects individuals' differences in concern with cooperation and social harmony. Agreeable individuals value getting along with others. They are therefore considerate, friendly, generous, helpful, and willing to compromise their interest with others. Agreeable people also have

an optimistic view of human nature. They believe people are basically honest, decent, and trustworthy.

Disagreeable individuals place self-interest above getting along with others. They are generally unconcerned with others' well-being, and therefore are unlikely to extend themselves to other people. Sometimes their skepticism about others' motives causes them to be suspicious, unfriendly, and uncooperative.

Agreeableness is obviously advantageous for attaining and maintaining popularity. Agreeable people are better liked than disagreeable people. On the other hand, agreeableness is not useful in situations that require tough or absolute objective decisions. Disagreeable people can make excellent scientists, critics, or soldiers.

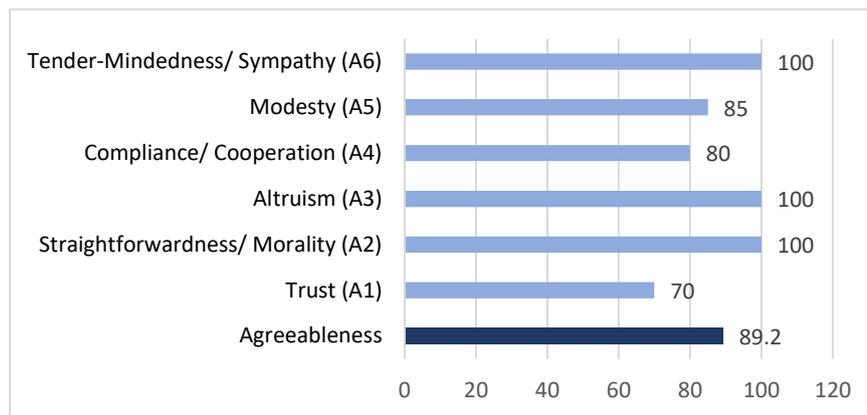

Your high level of agreeableness indicates a strong interest in others' needs and well-being. You are pleasant, sympathetic, and cooperative.

*Agreeableness Facets:*

- Trust (A1) – A person with high trust assumes that most people are fair, honest, and have good intentions. Persons low in the trust see others as selfish, devious, and potentially dangerous. Your level of trust is high.
- Morality (A2) – High scorers on this scale see no need for pretence or manipulation when dealing with others and are therefore candid, frank, and sincere. Low scorers believe that a certain amount of deception in social relationships is necessary. People find it relatively easy to relate to the straightforward high scorers on this scale. They generally find it more difficult to relate to the unstraightforward low scorers on this scale. It should be made clear that low scorers are not unprincipled or immoral; they are simply more guarded and less willing to openly reveal the whole truth. Your level of morality is high.
- Altruism (A3) – Altruistic people find helping other people genuinely rewarding. Consequently, they are generally willing to assist those who are in need. Altruistic people find that doing things for others is a form of self-fulfilment rather than self-sacrifice. Low scorers on this scale do not particularly like helping those in need. Requests for help feel like an imposition rather than an opportunity for self-fulfilment. Your level of altruism is high.
- Cooperation (A4) – Individuals who score high on this scale dislike confrontations. They are perfectly willing to compromise or deny their own needs in order to get along with others. Those who score low on this scale are more likely to intimidate others to get their way. Your level of cooperation is high.
- Modesty (A5) – High scorers on this scale do not like to claim that they are better than other people. In some cases, this attitude may derive from low self-confidence or self-esteem.

- Nonetheless, some people with high self-esteem find immodesty unseemly. Those who are willing to describe themselves as superior tend to be seen as disagreeably arrogant by other people. Your level of modesty is high.
- Sympathy (A6) – People who score high on this scale are tender-hearted and compassionate. They feel the pain of others, vicariously and are easily moved to pity. Low scorers are not affected strongly by human suffering. They pride themselves on making objective judgments based on reason. They are more concerned with truth and impartial justice than with mercy. Your level of tendermindedness is high.

**Conscientiousness**

Conscientiousness concerns the way in which we control, regulate, and direct our impulses. Impulses are not inherently bad; occasionally time constraints require a snap decision and acting on our first impulse can be an effective response. Also, in times of play rather than work, acting spontaneously and impulsively can be fun. Impulsive individuals can be seen by others as colourful, fun-to-be-with, and zany.

Nonetheless, acting on impulse can lead to trouble in a number of ways. Some impulses are antisocial. Uncontrolled antisocial acts not only harm other members of society but also can result in retribution toward the perpetrator of such impulsive acts. Another problem with impulsive acts is that they often produce immediate rewards but undesirable, long-term consequences. Examples include excessive socializing that leads to being fired from one's job, hurling an insult that causes the breakup of an important relationship or using pleasure-inducing drugs that eventually destroy one's health.

Impulsive behaviour, even when not seriously destructive, diminishes a person's effectiveness in significant ways. Acting impulsively disallows contemplating alternative courses of action, some of which would have been wiser than the impulsive choice. Impulsivity also side-tracks people during projects that require organized sequences of steps or stages. Accomplishments of an impulsive person are therefore small, scattered, and inconsistent.

A hallmark of intelligence, what potentially separates human beings from earlier life forms, is the ability to think about future consequences before acting on an impulse. Intelligent activity involves contemplation of long-range goals, organizing and planning routes to these goals, and persisting toward one's goals in the face of short-lived impulses to the contrary. The idea that intelligence involves impulse control is nicely captured by the term prudence, an alternative label for the Conscientiousness domain. Prudent means both wise and cautious. Persons who score high on the Conscientiousness scale are, in fact, perceived by others as intelligent.

The benefits of high conscientiousness are obvious. Conscientious individuals avoid trouble and achieve high levels of success through purposeful planning and persistence. They are also positively regarded by others as intelligent and reliable. On the negative side, they can be compulsive perfectionists and workaholics. Furthermore, extremely conscientious individuals might be regarded as stuffy and boring. Unconscientious people may be criticized for their unreliability, lack of ambition, and failure to stay within the lines, but they will experience many short-lived pleasures and they will never be called stuffy.

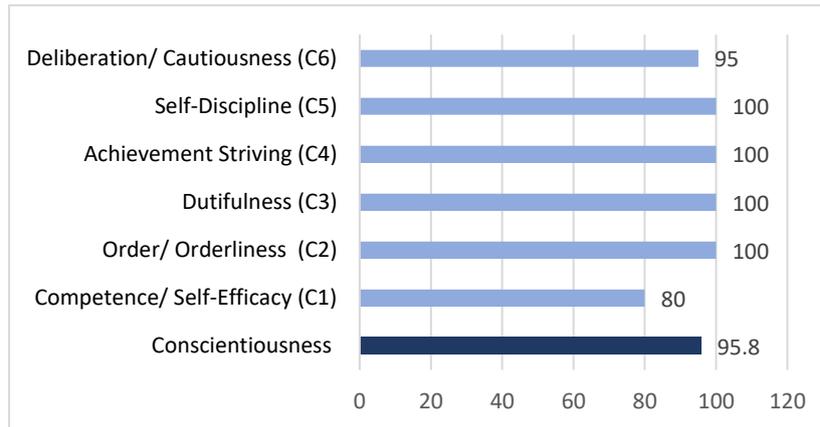

Your score on conscientiousness is high. This means you set clear goals and pursue them with determination. People regard you as reliable and hard-working.

*Conscientiousness Facets:*

- Self-Efficacy (C1) – Self-Efficacy describes confidence in one's ability to accomplish things. High scorers believe they have the intelligence (common sense), drive, and self-control necessary for achieving success. Low scorers do not feel effective and may have a sense that they are not in control of their lives. Your level of self-efficacy is high.
- Orderliness (C2) – Persons with high scores on orderliness are well-organized. They like to live according to routines and schedules. They keep lists and make plans. Low scorers tend to be disorganized and scattered. Your level of orderliness is high.
- Dutifulness (C3) – This scale reflects the strength of a person's sense of duty and obligation. Those who score high on this scale have a strong sense of moral obligation. Low scorers find contracts, rules, and regulations overly confining. They are likely to be seen as unreliable or even irresponsible. Your level of dutifulness is high.
- Achievement-Striving (C4) – Individuals who score high on this scale strive hard to achieve excellence. Their drive to be recognized as successful keeps them on track toward their lofty goals. They often have a strong sense of direction in life, but extremely high scores may be too single-minded and obsessed with their work. Low scorers are content to get by with a minimal amount of work and might be seen by others as lazy. Your level of achievement striving is high.
- Self-Discipline (C5) – Self-discipline-what many people call will-power-refers to the ability to persist at difficult or unpleasant tasks until they are completed. People who possess high self-discipline are able to overcome reluctance to begin tasks and stay on track despite distractions. Those with low self-discipline procrastinate and show poor follow-through, often failing to complete tasks-even tasks they want very much to complete. Your level of self-discipline is high.
- Cautiousness (C6) – Cautiousness describes the disposition to think through possibilities before acting. High scorers on the Cautiousness scale take their time when making decisions. Low scorers often say or do the first thing that comes to mind without deliberating alternatives and the probable consequences of those alternatives. Your level of cautiousness is high.

**Neuroticism**

Neuroticism refers to the tendency to experience negative feelings. Those who score high on Neuroticism may experience primarily one specific negative feeling such as anxiety, anger, or depression., but are likely to experience several of these emotions. People high in neuroticism are emotionally reactive. They respond emotionally to events that would not affect most people, and their reactions tend to be more intense than normal. They are more likely to interpret ordinary situations as threatening, and minor frustrations as hopelessly difficult. Their negative emotional reactions tend to persist for unusually long periods of time, which means they are often in a bad mood. These problems in emotional regulation can diminish a neurotic's ability to think clearly, make decisions, and cope effectively with stress.

At the other end of the scale, individuals who score low in neuroticism are less easily upset and are less emotionally reactive. They tend to be calm, emotionally stable, and free from persistent negative feelings. Freedom from negative feelings does not mean that low scorers experience a lot of positive feelings; the frequency of positive emotions is a component of the Extraversion domain.

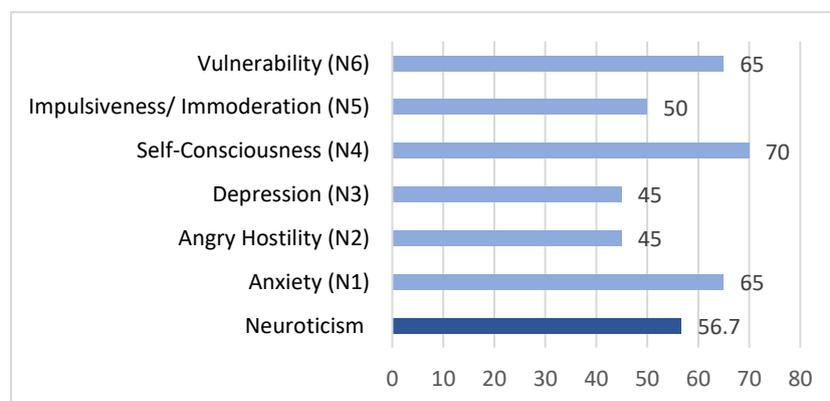

Your score on neuroticism is average, indicating that your level of emotional reactivity is typical of the general population. Stressful and frustrating situations are somewhat upsetting to you, but you are generally able to get over these feelings and cope with these situations.

*Neuroticism Facets:*

- Anxiety (N1) – The "fight-or-flight" system of the brain of anxious individuals is too easily and too often engaged. Therefore, people who are high in anxiety often feel like something dangerous is about to happen. They may be afraid of specific situations or be just generally fearful. They feel tense, jittery, and nervous. Persons low in Anxiety are generally calm and fearless. Your level of anxiety is average.
- Anger (N2) – Persons who score high in Anger feel enraged when things do not go their way. They are sensitive about being treated fairly and feel resentful and bitter when they feel they are being cheated. This scale measures the tendency to feel angry; whether or not the person expresses annoyance and hostility depends on the individual's level of Agreeableness. Low scorers do not get angry often or easily. Your level of anger is average.
- Depression (N3) – This scale measures the tendency to feel sad, dejected, and discouraged. High scorers lack energy and have difficulty initiating activities. Low scorers tend to be free from these depressive feelings. Your level of depression is average.
- Self-Consciousness (N4) – Self-conscious individuals are sensitive about what others think of them. Their concern about rejection and ridicule causes them to feel shy and uncomfortable around others. They are easily embarrassed and often feel ashamed. Their fears that others

will criticize or make fun of them are exaggerated and unrealistic, but their awkwardness and discomfort may make these fears a self-fulfilling prophecy. Low scorers, in contrast, do not suffer from the mistaken impression that everyone is watching and judging them. They do not feel nervous in social situations. Your level of self-consciousness is high.
- Immoderation (N5) – Immoderate individuals feel strong cravings and urge that they have difficulty resisting. They tend to be oriented toward short-term pleasures and rewards rather than long-term consequences. Low scorers do not experience strong, irresistible cravings and consequently do not find themselves tempted to overindulge. Your level of immoderation is average.
- Vulnerability (N6) – High scorers on Vulnerability experience panic, confusion, and helplessness when under pressure or stress. Low scorers feel more poised, confident, and clear-thinking when stressed. Your level of vulnerability is average.

**Openness to Experience**

Openness to Experience describes a dimension of cognitive style that distinguishes imaginative, creative people from down-to-earth, conventional people. Open people are intellectually curious, appreciative of art and sensitive to beauty. They tend to be, compared to closed people, more aware of their feelings. They tend to think and act in individualistic and nonconforming ways. Intellectuals typically score high on Openness to Experience; consequently, this factor has also been called Culture or Intellect. Nonetheless, Intellect is probably best regarded as one aspect of openness to experience. Scores on Openness to Experience are only modestly related to years of education and scores on standard intelligent tests.

Another characteristic of the open cognitive style is a facility for thinking in symbols and abstractions far removed from concrete experience. Depending on the individual's specific intellectual abilities, this symbolic cognition may take the form of mathematical, logical, or geometric thinking, artistic and metaphorical use of language, music composition or performance, or one of the many visual or performing arts. People with low scores on openness to experience tend to have narrow, common interests. They prefer the plain, straightforward, and obvious over the complex., ambiguous, and subtle. They may regard the arts and sciences with suspicion regarding these endeavours as abstruse or of no practical use. Closed people prefer familiarity over novelty; they are conservative and resistant to change.

Openness is often presented as healthier or more mature by psychologists, who are often themselves open to experience. However, open, and closed styles of thinking are useful in different environments. The intellectual style of the open person may serve a professor well, but research has shown that closed thinking is related to superior job performance in police work, sales, and a number of service occupations.

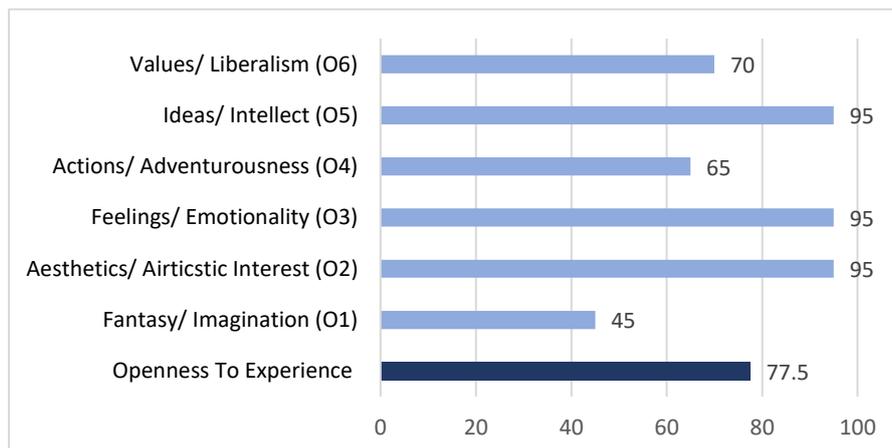

Your score on Openness to Experience is high, indicating you enjoy novelty, variety, and change. You are curious, imaginative, and creative.

***Openness Facets:***

- Imagination (O) – To imaginative individuals, the real world is often too plain and ordinary. High scorers on this scale use fantasy as a way of creating a richer, more interesting world. Low scorers on this scale are more oriented to facts than fantasy. Your level of imagination is average.
- Artistic Interests (O2) – High scorers on this scale love beauty. both in art and in nature. They become easily involved and absorbed in artistic and natural events. They are not necessarily artistically trained nor talented, although many will be. The defining features of this scale are interest in, and appreciation of natural and artificial beauty. Low scorers lack aesthetic sensitivity and interest in the arts. Your level of artistic interest is high.
- Emotionality (O3) – Persons high on Emotionality have good access to and awareness of their own feelings. Low scorers are less aware of their feelings and tend not to express their emotions openly. Your level of emotionality is high.
- Adventurousness (O4) – High scorers on adventurousness are eager to try new activities, travel to foreign lands, and experience different things. They find familiarity and routine boring and will take a new route home just because it is different. Low scorers tend to feel uncomfortable with change and prefer familiar routines. Your level of adventurousness is average.
- Intellect (O5) – Intellect and artistic interests are the two most important, central aspects of openness to experience. High scorers on Intellect love to play with ideas. They are open-minded to new and unusual ideas and like to debate intellectual issues. They enjoy riddles, puzzles, and brain teasers. Low scorers on Intellect prefer dealing with either people or things rather than ideas. They regard intellectual exercises as a waste of time. Intellect should not be equated with intelligence. Intellect is an intellectual style, not an intellectual ability, although high scorers on Intellect score slightly higher than low-Intellect individuals on standardized intelligence tests. Your level of intellect is high.
- Liberalism (O6) – Psychological liberalism refers to a readiness to challenge authority. convention, and traditional values. In its most extreme form. psychological liberalism can even represent outright hostility toward rules, sympathy for lawbreakers, and a love of ambiguity, chaos, and disorder. Psychological conservatives prefer the security and stability brought by conformity to tradition. Psychological liberalism and conservatism are not identical to political affiliation, but certainly, incline individuals toward certain political parties. Your level of liberalism is high.



\end{document}